\documentclass[a4paper,fleqn,usenatbib]{mnras}

\usepackage{graphicx}

\newcommand{\vsini}{\mbox{$v \sin i$}}

\newcommand{\etal}{et\,al.}
\newcommand{\kmps}{km\,s$^{-1}$}

\newcommand{\Msolar}{\mbox{M$_{\sun}$}}

\newcommand{\sqiglt}{\hbox{\rlap{\lower.55ex \hbox {$\sim$}}
	\kern-.3em \raise.4ex \hbox{$<$}\,}}
\newcommand{\sqiggt}{\hbox{\rlap{\lower.55ex \hbox {$\sim$}}
	\kern-.3em \raise.4ex \hbox{$>$}\,}}

\title[WASP-South transiting exoplanets]{WASP-South transiting exoplanets: WASP-130b, WASP-131b, WASP-132b, WASP-139b, WASP-140b, WASP-141b \&\ WASP-142b}

\author[Hellier et al.]{Coel Hellier$^{1}$,  
D.R. Anderson$^{1}$, 
A. Collier Cameron$^{2}$, 
L. Delrez$^{3}$,
M. Gillon$^{3}$,\newauthor  
E. Jehin$^{3}$, 
M. Lendl$^{4,5}$, 
P.F.L. Maxted$^{1}$, 
M. Neveu-VanMalle$^{5,6}$,
F. Pepe$^{5}$, \newauthor  
D. Pollacco$^{7}$, 
D. Queloz$^{6}$, 
D. S\'egransan$^{5}$, 
B. Smalley$^{1}$, 
J. Southworth$^{1}$, \newauthor  
A.H.M.J. Triaud$^{5,8}$, 
S. Udry$^{5}$, 
T. Wagg$^{1}$ \&\   
R.G. West$^{7}$\\    
$^{1}$Astrophysics Group, Keele University, Staffordshire, ST5 5BG, UK\\
$^{2}$SUPA, School of Physics and Astronomy, University of St.\ Andrews, North Haugh,  Fife, KY16 9SS, UK\\
$^{3}$Institut d'Astrophysique et de G\'eophysique, Universit\'e de
Li\`ege, All\'ee du 6 Ao\^ut, 17, Bat. B5C, Li\`ege 1, Belgium\\
$^{4}$Space Research Institute, Austrian Academy of Sciences, Schmiedlstr. 6, 8042, Graz, Austria\\
$^{5}$Observatoire astronomique de l'Universit\'e de Gen\`eve
51 ch. des Maillettes, 1290 Sauverny, Switzerland\\
$^{6}$Cavendish Laboratory, J J Thomson Avenue, Cambridge, CB3 0HE, UK\\
$^{7}$Department of Physics, University of Warwick, Gibbet Hill Road, Coventry, CV4 7AL, UK\\
$^{8}$Institute of Astronomy,  University of Cambridge,  Cambridge, CB3 0HA, UK\\
}

\begin{document}

\date{date}
\pagerange{range}

\maketitle

\begin{abstract}
We describe seven exoplanets transiting stars of brightness $V$ = 10.1 to 12.4. WASP-130b is a ``warm Jupiter'' having an orbital period of 11.6 d around a metal-rich G6 star. Its mass and radius  (1.23 $\pm$ 0.04 M$_{\rm Jup}$; 0.89 $\pm$ 0.03 R$_{\rm Jup}$) support the trend that warm Jupiters have smaller radii than hot Jupiters.  WASP-131b is a bloated Saturn-mass planet (0.27 M$_{\rm Jup}$; 1.22 R$_{\rm Jup}$). Its large scale height and bright ($V$ = 10.1) host star make it a good target for atmospheric characterisation.  WASP-132b  (0.41 M$_{\rm Jup}$; 0.87 R$_{\rm Jup}$) is among the least irradiated and coolest of WASP planets, having a 7.1-d orbit around a K4 star. WASP-139b is  a ``super-Neptune'' akin to HATS-7b and HATS-8b, being the lowest-mass planet yet found by WASP (0.12 M$_{\rm Jup}$; 0.80 R$_{\rm Jup}$). The metal-rich K0 host star appears to be anomalously dense, akin to HAT-P-11. WASP-140b is a 2.4-M$_{\rm Jup}$ planet in an eccentric ($e = 0.047 \pm 0.004$)  2.2-d orbit. The planet's radius is large (1.4 R$_{\rm Jup}$), but uncertain owing to the grazing transit ($b$ = 0.93). The 10.4-day rotation period of the K0 host star suggests a young age, and the timescale for tidal circularisation is likely to be the lowest of all known eccentric hot Jupiters.  WASP-141b (2.7 M$_{\rm Jup}$, 1.2 R$_{\rm Jup}$, $P$ = 3.3 d) and WASP-142b (0.84 M$_{\rm Jup}$, 1.53 R$_{\rm Jup}$,  $P$ = 2.1 d) are typical hot Jupiters orbiting metal-rich F stars. We show that the period distribution within the hot-Jupiter bulge does not depend on the metallicity of the host star. 
\end{abstract}


\section{Introduction}
The WASP survey continues to be a productive means of finding giant planets transiting relatively bright stars. WASP discoveries are often prime targets for further study. For example \citet{2016Natur.529...59S} devoted 125 orbits of {\it Hubble Space Telescope\/} time to exoplanet atmospheres, of which 6 out of 8 targets were WASP planets.  Similarly,   \citet{2016PASP..128i4401S} propose 12 planets as ``community targets'' for atmospheric characterisation in Cycle 1 of the {\it James Webb Space Telescope}, of which 7 are WASP planets.  Ongoing discoveries also increase the census of closely orbiting giant planets, and continue to find planets with novel characteristics.   

Here we report seven new transiting giant planets discovered by the WASP-South survey instrument in conjunction with the Euler/CORALIE spectrograph and the robotic  TRAPPIST photometer.   With 200-mm lenses the eight WASP-South cameras can cover up to half the available sky per year (south of declination +08$^{\circ}$ and avoiding the crowded galactic plane). This means that the data that led to the current discoveries, accumulated from 2006 May to 2012 Jun, typically includes three seasons of coverage, or more where pointings overlap.  Combining multiple years of observation  gives sensitivity to longer orbital periods, and this batch of planets includes the longest-period WASP discovery yet, a ``warm Jupiter'' at 11.6 days.

\begin{table}
\caption{Observations\protect\rule[-1.5mm]{0mm}{2mm}}  
\begin{tabular}{lcr}
\hline 
Facility & Date & Notes \\ [0.5mm] \hline
\multicolumn{3}{l}{{\bf WASP-130:}}\\  
WASP-South & 2006 May--2012 Jun & 28\,400 points \\ 
CORALIE  & 2014 Feb--2016 Mar  &   27 RVs \\
EulerCAM  & 2014 May 04 & Gunn $r$ filter \\ 
EulerCAM  & 2014 May 27 & Gunn $r$ filter \\ 
TRAPPIST & 2014 May 27 & $z^{\prime}$ band \\
TRAPPIST & 2015 Feb 05 & $z^{\prime}$ band \\
TRAPPIST & 2015 Apr 27 & $z^{\prime}$ band \\
EulerCAM  & 2015 Apr 27 & Gunn $r$ filter \\ 
\multicolumn{3}{l}{{\bf WASP-131:}}\\  
WASP-South & 2006 May--2012 Jun & 23\,300 points \\ 
CORALIE  & 2014 Feb--2016 Mar  &   23 RVs \\
TRAPPIST & 2014 Apr 22 & $z$ band \\
EulerCAM  & 2014 Apr 22 & Gunn $r$ filter \\ 
EulerCAM  & 2015 Mar 02 & $I_{c}$ filter \\ 
EulerCAM  & 2015 Apr 03 & $I_{c}$ filter \\ 
TRAPPIST & 2015 Apr 19 & $z$ band \\
TRAPPIST & 2015 Jun 06 & $z^{\prime}$ band \\
\multicolumn{3}{l}{{\bf WASP-132:}}\\  
WASP-South & 2006 May--2012 Jun & 23\,300 points \\ 
CORALIE  & 2014 Mar--2016 Mar  &   36 RVs \\
TRAPPIST & 2014 May 05 & $I+z$ band \\
\multicolumn{3}{l}{{\bf WASP-139:}}\\  
WASP-South & 2006 May--2012 Jun & 21\,000 points \\ 
CORALIE  & 2008 Oct--2015 Dec  &   24 RVs \\
HARPS    & 2014 Sep--2015 Jan  &   27 RVs \\
TRAPPIST & 2014 Aug 06 & $I+z$ band \\
TRAPPIST & 2015 Sep 07 & $I+z$ band \\
EulerCAM & 2015 Sep 07 & {\it NGTS\/} filter \\ 
\multicolumn{3}{l}{{\bf WASP-140:}}\\  
WASP-South & 2006 Aug--2012 Jan & 31\,300 points \\ 
CORALIE  & 2014 Sep--2015 Dec  &   23 RVs \\
TRAPPIST & 2014 Oct 12 & $z^{\prime}$ band \\
TRAPPIST & 2014 Nov 28 & $z^{\prime}$ band \\
TRAPPIST & 2014 Dec 07 & $z^{\prime}$ band \\
EulerCAM & 2015 Sep 01 & $I_{c}$ filter \\ 
\multicolumn{3}{l}{{\bf WASP-141:}}\\  
WASP-South & 2006 Sep--2012 Feb & 21\,400 points \\ 
CORALIE  & 2014 Oct--2015 Dec  &   18 RVs \\
EulerCAM & 2014 Dec 17 & {\it NGTS\/} filter \\ 
TRAPPIST & 2015 Jan 19 & $I+z$ band \\
TRAPPIST & 2015 Dec 26 & $I+z$ band \\
\multicolumn{3}{l}{{\bf WASP-142:}}\\  
WASP-South & 2006 May--2012 May & 47\,300 points \\ 
CORALIE  & 2014 Oct--2016 Feb  &   16 RVs \\
TRAPPIST & 2014 May 26 & Blue-block \\
EulerCAM  & 2014 Dec 13 & $I_{c}$ filter \\ 
EulerCAM  & 2015 Mar 01 & $I_{c}$ filter \\ 
TRAPPIST & 2015 Apr 03 & Blue-block \\
\end{tabular} 
\end{table} 

\section{Observations}
Since the processes and techniques used here are a continuation of those from other recent WASP-South discovery papers  (e.g.\ \citealt{2014MNRAS.445.1114A}; \citealt{2014MNRAS.440.1982H}; \citealt{2016A&A...591A..55M}) we describe them briefly. The WASP camera arrays \citep{2006PASP..118.1407P} tile fields of $7.8^{\circ}\times7.8^{\circ}$ with a typical cadence of 10 mins, using 200mm f/1.8 lenses backed by 2k$\times$2k Peltier-cooled CCDs.  Using transit-search algorithms  \citep{2007MNRAS.380.1230C}  we trawl the accumulated multi-year lightcurves for planet candidates, which are then passed to the  1.2-m Euler/CORALIE spectrograph (e.g. \citealt{2013A&A...551A..80T}), for radial-velocity observations, and to the robotic 0.6-m TRAPPIST photometer,  which resolves candidates which are blended in WASP's large, 14$^{\prime\prime}$, pixels.  TRAPPIST (e.g. \citealt{2013A&A...552A..82G}) and EulerCAM (e.g. \citealt{2012A&A...544A..72L}) then obtain higher-quality photometry of newly confirmed planets.    For one system reported here, WASP-139, we have also obtained radial velocities using the HARPS spectrometer on the ESO 3.6-m (\citealt{2003Msngr.114...20M}).   A list of our observations is given in Table~1 while the radial velocities are listed in Table~A1.

\section{The host stars} 
We used the CORALIE spectra to estimate spectral parameters of the host stars using the  methods described in \citet{2013MNRAS.428.3164D}.  We used the H$\alpha$ line to estimate the effective temperature ($T_{\rm eff}$), and the Na~{\sc i} D and Mg~{\sc i} b lines as diagnostics of the surface gravity ($\log g$). The Iron abundances were determined from equivalent-width measurements of several clean and unblended Fe~{\sc i} lines and are given relative to the Solar value presented in \citet{2009ARA&A..47..481A}. The quoted abundance errors include that given by the uncertainties in $T_{\rm eff}$ and $\log g$, as well as the scatter due to measurement and atomic data uncertainties. The projected rotation velocities ($v \sin i$) were determined by fitting the profiles of the Fe~{\sc i} lines after convolving with the CORALIE instrumental resolution ($R$ = 55\,000) and a macroturbulent velocity adopted from the calibration of \citet{2014MNRAS.444.3592D}. 

The parameters obtained from the analysis are given in Tables 2 to 8. Gyrochronological age estimates are given for three stars, derived from the
measured \vsini\ and compared to values in \citet{2007ApJ...669.1167B}; for the other stars no sensible constraint is obtained.  Lithium age estimates come
from values in  \citet{2005A&A...442..615S}. We also list proper motions
from the UCAC4 catalogue \citep{2013AJ....145...44Z}.

We searched the WASP photometry of each star for rotational
modulations by using a sine-wave fitting algorithm as described by \citet{2011PASP..123..547M}. We estimated the significance of periodicities
by subtracting the fitted transit lightcurve and then repeatedly and
randomly permuting the nights of observation.  We found a significant 
modulation in WASP-140 (see Section 10) and a possible modulation in WASP-132 (Section 8) and report upper limits for the other stars. 

\section{System parameters}
The CORALIE radial-velocity measurements (and the HARPS data for
WASP-139) were combined with the WASP, EulerCAM and TRAPPIST
photometry in a simultaneous Markov-chain Monte-Carlo (MCMC) analysis
to find the system parameters. CORALIE was upgraded in 2014 November,
and so we treat the RV data before and after that time as independent
datasets, allowing a zero-point offset between them (the division is
indicated by a short horizontal line in Table~A1).  For more details
of our methods see \citet{2007MNRAS.375..951C}.  The limb-darkening
parameters are noted in each Table, and are taken from the 4-parameter
non-linear law of \citet{2000A&A...363.1081C}.

For WASP-140b the orbital eccentricity is significant and was fitted
as a free parameter.  For the others we imposed a circular orbit since
hot Jupiters are expected to circularise on a timescale less than
their age, and so adopting a circular orbit gives the most likely
parameters (see, e.g., \citealt{2012MNRAS.422.1988A}).

The fitted parameters were $T_{\rm c}$, $P$, $\Delta F$, $T_{14}$,
$b$, $K_{\rm 1}$, where $T_{\rm c}$ is the epoch of mid-transit, $P$
is the orbital period, $\Delta F$ is the fractional flux-deficit that
would be observed during transit in the absence of limb-darkening,
$T_{14}$ is the total transit duration (from first to fourth contact),
$b$ is the impact parameter of the planet's path across the stellar
disc, and $K_{\rm 1}$ is the stellar reflex velocity
semi-amplitude. 

The transit lightcurves lead directly to stellar density but one
additional constraint is required to obtain stellar masses and radii,
and hence full parametrisation of the system. As with other recent
WASP discovery papers, we compare the derived stellar density and the
spectroscopic effective temperature and metallicity to a grid of
stellar models, as described in \citet{2015A&A...575A..36M}.  We use
an MCMC method to calculate the posterior distribution for the mass
and age estimates of the star.  The stellar models were calculated
using the {\sc garstec} stellar evolution code
\citep{2008Ap&SS.316...99W} and the methods used to calculate the
stellar model grid are described in \citet{2013MNRAS.429.3645S}.

\begin{figure}
\hspace*{2mm}\includegraphics[width=8.5cm]{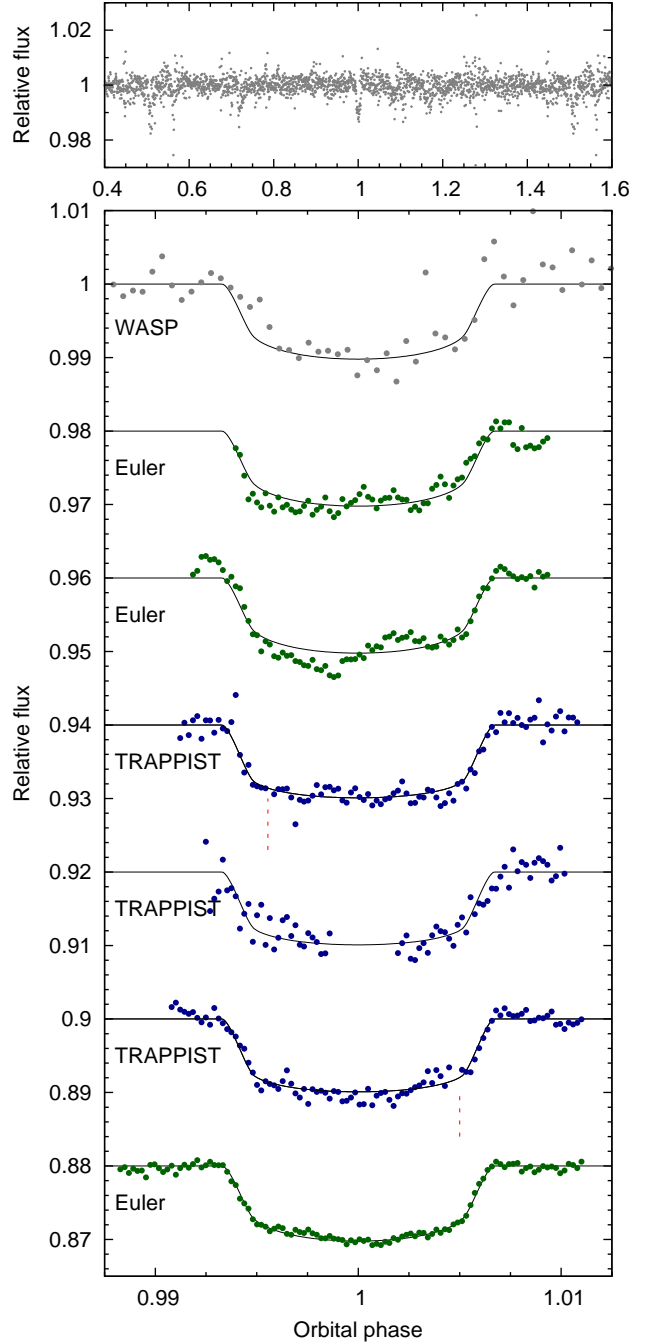}\\ [-2mm]
\caption{WASP-130b discovery photometry: (Top) The WASP data folded on the 
transit period. (Second panel) The binned WASP data with (offset) the
follow-up transit lightcurves (ordered from the top as in Table~1) together with the fitted MCMC model. Red dashed lines indicate times when the TRAPPIST photometer was flipped across the meridian. The second EulerCAM lightcurve had poor observing conditions and shows excess red noise, which was accounted for by inflating the errors in the MCMC process.}
\end{figure}

\begin{figure}
\hspace*{2mm}\includegraphics[width=8.5cm]{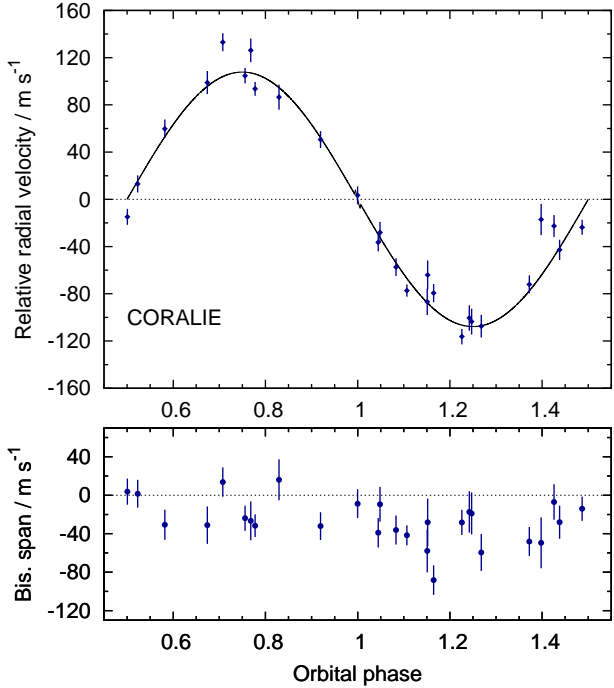}\\ [-2mm]
\caption{WASP-130b radial velocities and fitted model (top) along with (bottom) the bisector spans; the absence of any correlation with radial velocity is a check against transit mimics.}
\end{figure}

For each system we list the resulting parameters in Tables~2 to 8 and
show the data and models in Figures 1 to 14.  We generally report
1-$\sigma$ error bars on all quantities.  For the possible effects of
red noise in transit lightcurves and their affect on system parameters
see the extensive analysis by \citet{2012AJ....143...81S}. We report
the comparison to stellar models in Table~\ref{AgeMassTable}, where we give the likeliest age and the 95\%\ confidence interval,  and display the comparison in
Fig.~\ref{AgeMassFig}.

\begin{table}
\caption{System parameters for WASP-130.}  
\begin{tabular}{lc}
\multicolumn{2}{l}{1SWASP\,J133225.42--422831.0}\\
\multicolumn{2}{l}{2MASS\,13322543--4228309}\\
\multicolumn{2}{l}{RA\,=\,13$^{\rm h}$32$^{\rm m}$25.43$^{\rm s}$, 
Dec\,=\,--42$^{\circ}$28$^{'}$30.9$^{''}$ (J2000)}\\
\multicolumn{2}{l}{$V$ mag = 11.1}  \\ 
\multicolumn{2}{l}{Rotational modulation\ \ \ $<$\,1 mmag (95\%)}\\
\multicolumn{2}{l}{pm (RA) 7.0\,$\pm$\,2.5 (Dec) --0.7\,$\pm$\,1.3 mas/yr}\\
\hline
\multicolumn{2}{l}{Stellar parameters from spectroscopic analysis.\rule[-1.5mm]{
0mm}{2mm}} \\ \hline 
Spectral type & G6 \\
$T_{\rm eff}$ (K)  & 5600  $\pm$ 100  \\
$\log g$      & 4.4 $\pm$ 0.1    \\
$v\,\sin i$ (km\,s$^{-1}$)     &    0.5 $\pm$ 0.5     \\
{[Fe/H]}   &   +0.26 $\pm$ 0.10     \\
log A(Li)  &    $<$ 0.4      \\
Age (Lithium) [Gy]  &   $\ga$ 2         \\
Distance [pc]  &  180 $\pm$  30  \\ \hline 
\multicolumn{2}{l}{Parameters from MCMC analysis.\rule[-1.5mm]{0mm}{3mm}} \\
\hline 
$P$ (d) & 11.55098 $\pm$ 0.00001 \\
$T_{\rm c}$ (HJD)\,(UTC) & 245\,6921.14307 $\pm$ 0.00025 \\
$T_{\rm 14}$ (d) & 0.155 $\pm$ 0.001 \\
$T_{\rm 12}=T_{\rm 34}$ (d) & 0.018 $\pm$ 0.001 \\
$\Delta F=R_{\rm P}^{2}$/R$_{*}^{2}$ & 0.00916 $\pm$ 0.00014 \\
$b$ & 0.53 $\pm$ 0.03 \\
$i$ ($^\circ$)  & 88.66 $\pm$ 0.12 \\
$K_{\rm 1}$ (km s$^{-1}$) & 0.108 $\pm$ 0.002 \\
$\gamma$ (km s$^{-1}$)  & 1.462 $\pm$ 0.002 \\
$e$ & 0 (adopted) ($<$\,0.04 at 2$\sigma$) \\ 
$M_{\rm *}$ (M$_{\rm \odot}$) & 1.04 $\pm$ 0.04 \\
$R_{\rm *}$ (R$_{\rm \odot}$) & 0.96 $\pm$ 0.03 \\
$\log g_{*}$ (cgs) & 4.49 $\pm$ 0.02 \\
$\rho_{\rm *}$ ($\rho_{\rm \odot}$) & 1.18 $\pm$ 0.09\\
$T_{\rm eff}$ (K) & 5625 $\pm$ 90 \\
$M_{\rm P}$ (M$_{\rm Jup}$) & 1.23 $\pm$ 0.04 \\
$R_{\rm P}$ (R$_{\rm Jup}$) & 0.89 $\pm$ 0.03 \\
$\log g_{\rm P}$ (cgs) & 3.55 $\pm$ 0.03 \\
$\rho_{\rm P}$ ($\rho_{\rm J}$) & 1.76 $\pm$ 0.18 \\
$a$ (AU)  & 0.1012 $\pm$ 0.0014 \\
$T_{\rm P, A=0}$ (K) & 833 $\pm$ 18 \\ [0.5mm] \hline 
\multicolumn{2}{l}{Errors are 1$\sigma$; Limb-darkening coefficients were:}\\
\multicolumn{2}{l}{{\small $r$ band: a1 = 0.669, a2 = --0.400, a3 = 0.976, 
a4 = --0.481}}\\ 
\multicolumn{2}{l}{{\small $z$ band: a1 =   0.657 , a2 = --0.454 , a3 = 0.834, a4 = --0.407}}\\ \hline
\end{tabular} 
\end{table}

\section{WASP-130}
WASP-130 is a $V$ = 11.1, G6 star with a metallicity of [Fe/H] = +0.26
$\pm$ 0.10.  The transit $\log g_{*}$ of 4.49 $\pm$ 0.02 is consistent
with the spectroscopic $\log g_{*}$ of 4.4 $\pm$ 0.1.  The
evolutionary comparison (Fig.~\ref{AgeMassFig}) suggests an age of
0.2--7.9 Gyr (consistent with the lithium age estimate of $\ga$ 2
Gyr). 

The radial velocities show excess scatter with could be due to
magnetic activity, though in this system there is no detection of a rotational
modulation in the WASP data. Scatter when folded on the orbital period can also be caused by a longer-term trend, but that is not the case here. 

The planet, WASP-130b, has an orbital period of 11.6 days, the longest yet found by WASP-South, and is thus a ``warm jupiter''.  For comparison, the HATNet and HATSouth projects have cameras at more than one longitude and so are more sensitive to longer periods; their longest-period system is currently HATS-17b at 16.3 d \citep{2016AJ....151...89B}.

The mass of WASP-130b is 1.23 $\pm$ 0.04 M$_{\rm Jup}$.  In keeping with other longer-period systems (e.g.~\citealt{2011ApJS..197...12D}), but in contrast to many hotter Jupiters, the radius is not bloated (0.89 $\pm$ 0.03 R$_{\rm Jup}$).    WASP-130b is thus similar to  HATS-17b (1.34 M$_{\rm Jup}$; 0.78 R$_{\rm Jup}$; \citealt{2016AJ....151...89B}), though not quite as compact.   Brahm \etal\ suggest that HATS-17b has a massive metallic core, which they link to the raised metallicity of [Fe/H] = +0.3, which is again similar to that of WASP-130 ([Fe/H] = +0.25).

\begin{figure}
\hspace*{2mm}\includegraphics[width=8.5cm]{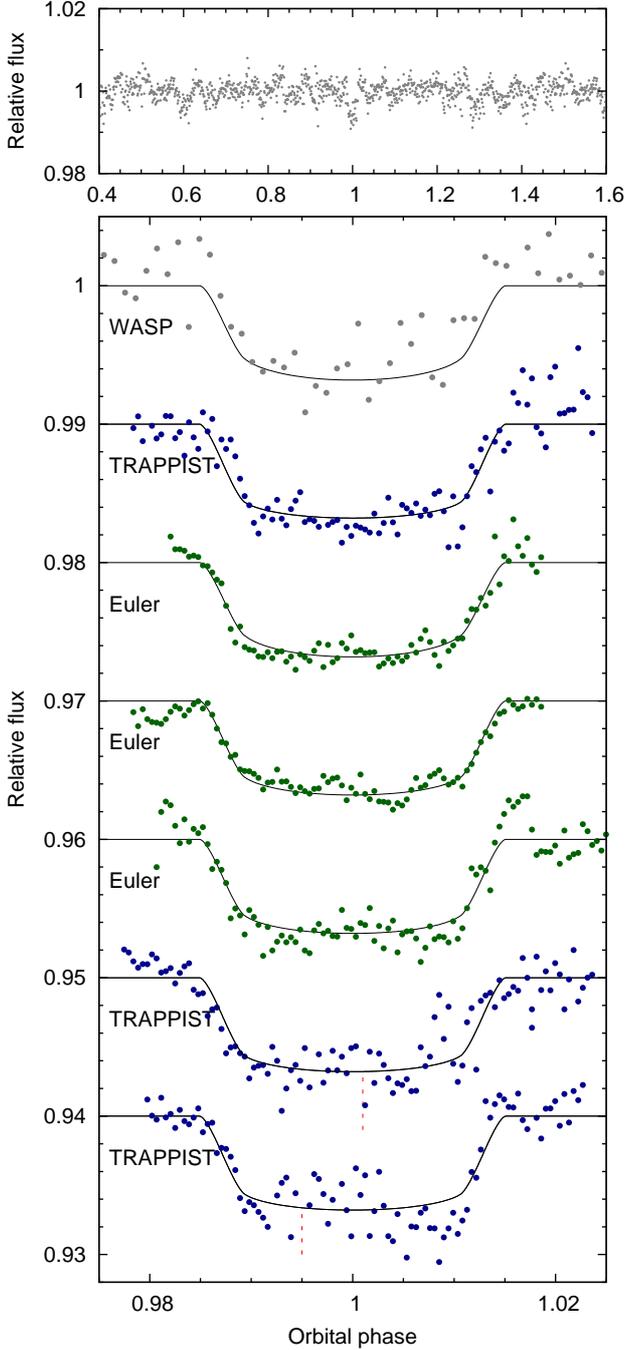}\\ [-2mm]
\caption{WASP-131b discovery photometry, as for Fig.~1.}
\end{figure}

\begin{figure}
\hspace*{2mm}\includegraphics[width=8.5cm]{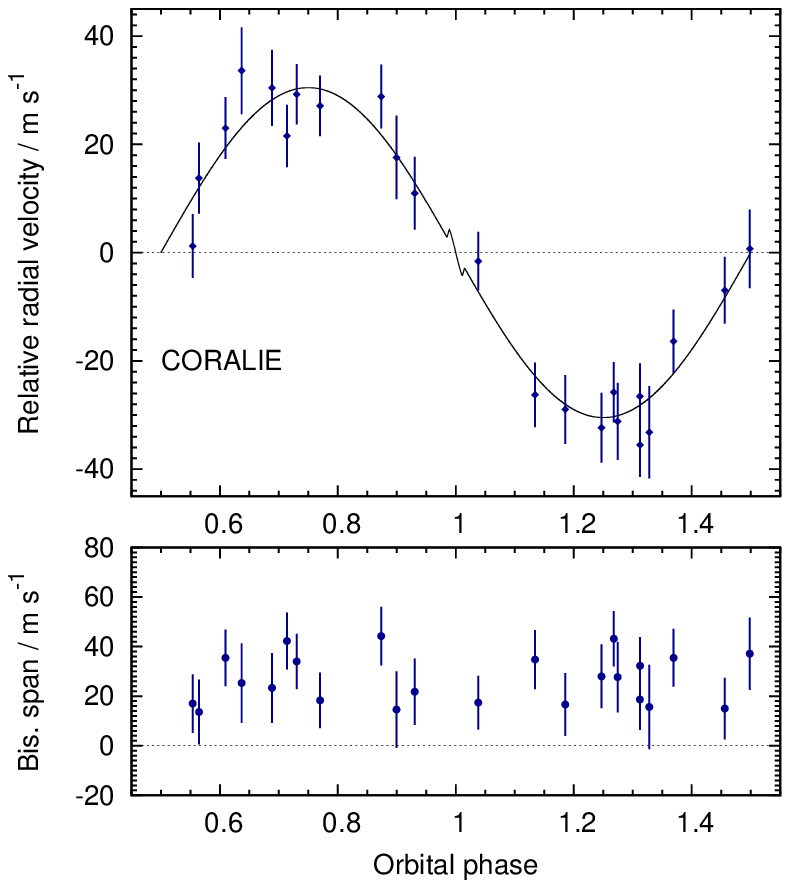}\\ [-2mm]
\caption{WASP-131b radial velocities and bisector spans, as for Fig.~2.}
\end{figure}

\begin{table}
\caption{System parameters for WASP-131.}  
\begin{tabular}{lc}
\multicolumn{2}{l}{1SWASP\,J140046.44--303500.8}\\
\multicolumn{2}{l}{2MASS\,14004645--3035008}\\
\multicolumn{2}{l}{RA\,=\,14$^{\rm h}$00$^{\rm m}$46.45$^{\rm s}$, 
Dec\,=\,--30$^{\circ}$35$^{'}$00.8$^{''}$ (J2000)}\\
\multicolumn{2}{l}{$V$ mag = 10.1}  \\ 
\multicolumn{2}{l}{Rotational modulation\ \ \ $<$\,0.5 mmag (95\%)}\\
\multicolumn{2}{l}{pm (RA) 11.4\,$\pm$\,1.9 (Dec) --6.5\,$\pm$\,1.0 mas/yr}\\
\hline
\multicolumn{2}{l}{Stellar parameters from spectroscopic analysis.\rule[-1.5mm]{
0mm}{2mm}} \\ \hline 
Spectral type & G0 \\
$T_{\rm eff}$ (K)  & 5950  $\pm$ 100  \\
$\log g$      & 3.9 $\pm$ 0.1    \\
$v\,\sin i$ (km\,s$^{-1}$)     &    3.0 $\pm$ 0.9     \\
{[Fe/H]}   &   --0.18 $\pm$ 0.08     \\
log A(Li)  &    2.60 $\pm$ 0.08   \\
Age (Lithium) [Gy]  &       1 $\sim$ 8       \\
Distance [pc]  &  250 $\pm$ 50   \\ \hline 
\multicolumn{2}{l}{Parameters from MCMC analysis.\rule[-1.5mm]{0mm}{3mm}} \\
\hline 
$P$ (d) & 5.322023 $\pm$ 0.000005 \\
$T_{\rm c}$ (HJD)\,(UTC) & 2456919.8236 $\pm$ 0.0004 \\
$T_{\rm 14}$ (d) & 0.1596 $\pm$ 0.0014 \\
$T_{\rm 12}=T_{\rm 34}$ (d) & 0.0243 $\pm$ 0.0016 \\
$\Delta F=R_{\rm P}^{2}$/R$_{*}^{2}$ & 0.00665 $\pm$ 0.00012 \\
$b$ & 0.73 $\pm$ 0.02 \\
$i$ ($^\circ$)  & 85.0 $\pm$ 0.3 \\
$K_{\rm 1}$ (km s$^{-1}$) & 0.0305 $\pm$ 0.0017 \\
$\gamma$ (km s$^{-1}$)  & --19.6636 $\pm$ 0.0015 \\
$e$ & 0 (adopted) ($<$\,0.10 at 2$\sigma$) \\ 
$M_{\rm *}$ (M$_{\rm \odot}$) & 1.06 $\pm$ 0.06 \\
$R_{\rm *}$ (R$_{\rm \odot}$) & 1.53 $\pm$ 0.05 \\
$\log g_{*}$ (cgs) & 4.089 $\pm$ 0.026 \\
$\rho_{\rm *}$ ($\rho_{\rm \odot}$) & 0.292 $\pm$ 0.026 \\
$T_{\rm eff}$ (K) & 6030 $\pm$ 90 \\
$M_{\rm P}$ (M$_{\rm Jup}$) & 0.27 $\pm$ 0.02 \\
$R_{\rm P}$ (R$_{\rm Jup}$) & 1.22 $\pm$ 0.05 \\
$\log g_{\rm P}$ (cgs) & 2.62 $\pm$ 0.04 \\
$\rho_{\rm P}$ ($\rho_{\rm J}$) & 0.15 $\pm$ 0.02 \\
$a$ (AU)  & 0.0607 $\pm$ 0.0009 \\
$T_{\rm P, A=0}$ (K) & 1460 $\pm$ 30 \\ [0.5mm] \hline 
\multicolumn{2}{l}{Errors are 1$\sigma$; Limb-darkening coefficients were:}\\
\multicolumn{2}{l}{{\small $r$ band: a1 = 0.601, a2 = --0.085, a3 = 0.517, 
a4 = --0.300}}\\ 
\multicolumn{2}{l}{{\small $I$ band: a1 = 0.676, a2 = --0.353, a3 = 0.685, 
a4 = --0.347}}\\ 
\multicolumn{2}{l}{{\small $z$ band: a1 =   0.573 , a2 = --0.142 , a3 = 0.410, a4 = --0.241}}\\ \hline
\end{tabular} 
\end{table}

\section{WASP-131}
WASP-131 is a $V$ = 10.1, G0 star with a metallicity of [Fe/H] =
--0.18 $\pm$ 0.08.  The transit $\log g_{*}$ of 4.09 $\pm$ 0.03 is
consistent with the spectroscopic $\log g_{*}$ of 3.9 $\pm$ 0.1.  The
radius is inflated (1.53 R$_{\odot}$ for 1.06 M$_{\odot}$) and the
evolutionary comparison (Fig.~\ref{AgeMassFig}) suggests an age of 4.5--10 Gyr (consistent with the poorly constrained lithium estimate
of between 1 and 8 Gyr).

The planet, WASP-131b, has an orbital period of 5.3 days.  It is a Saturn-mass but bloated planet (0.27 M$_{\rm Jup}$; 1.22 R$_{\rm Jup}$).   The low density of the planet (0.15\,$\pm$\,0.02\,$\rho_{\rm J}$) and the consequent large scale-height of the atmosphere, coupled with the host-star magnitude of $V$ = 10.1, should make WASP-131b a good candidate for atmospheric characterisation.  

Low-density, Saturn-mass planets akin to WASP-131b have been seen before.  The most similar include WASP-21b (0.28 M$_{\rm Jup}$; 1.2 R$_{\rm Jup}$; $P$ = 4.3 d; \citealt{2010A&A...519A..98B}), WASP-39b (0.28 M$_{\rm Jup}$; 1.3 R$_{\rm Jup}$; $P$ = 4.1 d; \citealt{2011A&A...531A..40F}), Kepler-427b (0.29 M$_{\rm Jup}$; 1.2 R$_{\rm Jup}$; $P$ = 10.3 d; \citealt{2014A&A...572A..93H}) and HAT-P-51b (0.30 M$_{\rm Jup}$; 1.3 R$_{\rm Jup}$; $P$ = 4.2 d; \citealt{2015AJ....150..168H}).


\begin{figure}
\hspace*{2mm}\includegraphics[width=8.5cm]{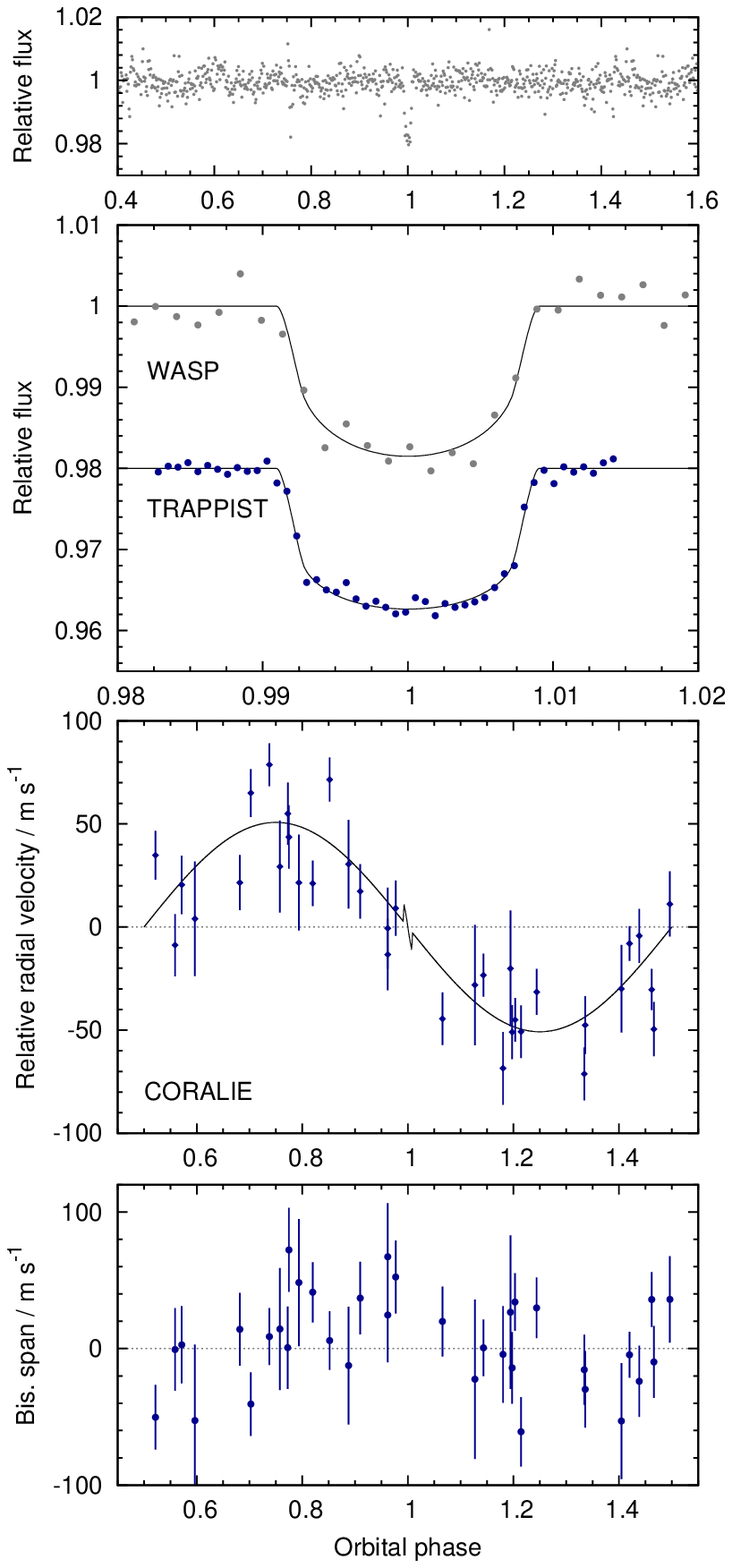}\\ [-2mm]
\caption{WASP-132b discovery data, as for Figs.~1 \&\ 2.}
\end{figure}

\begin{figure}
\hspace*{0mm}\includegraphics[width=9cm]{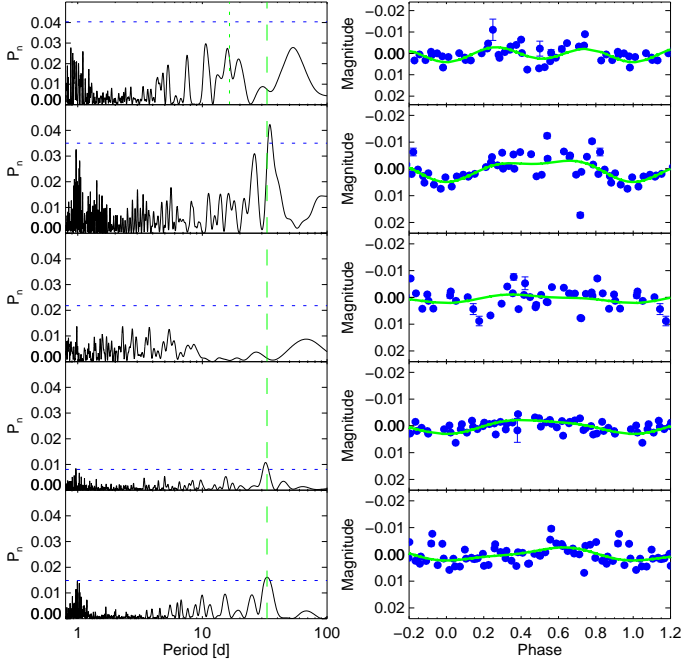}\\ [-2mm]
\caption{Possible rotational modulation in the WASP data for WASP-132.  The left-hand panels show peridograms for each season of data (2006, 2007, 2008, 2011 \&\ 2012 from the top down). The 33-d period is marked in green, as is half this period in the uppermost panel. The blue dotted line is a false-alarm probability of 0.001.  The right-hand panels show the data for each season folded on the 33.4-d period; the green line is a harmonic-series fit to the data}
\end{figure}

\begin{table}
\caption{System parameters for WASP-132.}  
\begin{tabular}{lc}
\multicolumn{2}{l}{1SWASP\,J143026.22--460933.0}\\
\multicolumn{2}{l}{2MASS\,14302619--4609330}\\
\multicolumn{2}{l}{RA\,=\,14$^{\rm h}$30$^{\rm m}$26.19$^{\rm s}$, 
Dec\,=\,--46$^{\circ}$09$^{'}$33.0$^{''}$ (J2000)}\\
\multicolumn{2}{l}{$V$ mag = 12.4}  \\ 
\multicolumn{2}{l}{Rotational modulation: possible 0.4-mmag at 33-d.}\\
\multicolumn{2}{l}{pm  (RA) 14.2\,$\pm$\,1.3 (Dec) --73.8\,$\pm$\,1.3 mas/yr}\\
\hline
\multicolumn{2}{l}{Stellar parameters from spectroscopic analysis.\rule[-1.5mm]{
0mm}{2mm}} \\ \hline 
Spectral type & K4 \\
$T_{\rm eff}$ (K)  & 4750  $\pm$ 100  \\
$\log g$      & 4.6 $\pm$ 0.1    \\
$v\,\sin i$ (km\,s$^{-1}$)     &    0.9 $\pm$ 0.8     \\
{[Fe/H]}   &   +0.22 $\pm$ 0.13     \\
log A(Li)  &    $<$ --0.3   \\
Age (Lithium) [Gy]  &   $\ga$ 0.5          \\
Distance [pc]  & 120 $\pm$ 20   \\ \hline 
\multicolumn{2}{l}{Parameters from MCMC analysis.\rule[-1.5mm]{0mm}{3mm}} \\
\hline 
$P$ (d) & 7.133521 $\pm$ 0.000009 \\
$T_{\rm c}$ (HJD)\,(UTC) & 2456698.2076 $\pm$ 0.0004 \\
$T_{\rm 14}$ (d) & 0.1284 $\pm$ 0.0009 \\
$T_{\rm 12}=T_{\rm 34}$ (d) & 0.0141 $\pm$ 0.0008 \\
$\Delta F=R_{\rm P}^{2}$/R$_{*}^{2}$ & 0.0146 $\pm$ 0.0003 \\
$b$ & 0.14 $\pm$ 0.12 \\
$i$ ($^\circ$)  & 89.6 $\pm$ 0.3 \\
$K_{\rm 1}$ (km s$^{-1}$) & 0.051 $\pm$ 0.003 \\
$\gamma$ (km s$^{-1}$)  & 31.067 $\pm$ 0.003 \\
$e$ & 0 (adopted) ($<$\,0.10 at 2$\sigma$) \\ 
$M_{\rm *}$ (M$_{\rm \odot}$) & 0.80 $\pm$ 0.04 \\
$R_{\rm *}$ (R$_{\rm \odot}$) & 0.74 $\pm$ 0.02 \\
$\log g_{*}$ (cgs) & 4.61 $\pm$ 0.02 \\
$\rho_{\rm *}$ ($\rho_{\rm \odot}$) & 2.00 $^{+0.07}_{-0.14}$ \\
$T_{\rm eff}$ (K) & 4775 $\pm$ 100 \\
$M_{\rm P}$ (M$_{\rm Jup}$) & 0.41 $\pm$ 0.03 \\
$R_{\rm P}$ (R$_{\rm Jup}$) & 0.87 $\pm$ 0.03 \\
$\log g_{\rm P}$ (cgs) & 3.10 $\pm$ 0.04 \\
$\rho_{\rm P}$ ($\rho_{\rm J}$) & 0.63 $\pm$ 0.06 \\
$a$ (AU)  & 0.067 $\pm$ 0.001 \\
$T_{\rm P, A=0}$ (K) & 763 $\pm$ 16 \\ [0.5mm] \hline 
\multicolumn{2}{l}{Errors are 1$\sigma$; Limb-darkening coefficients were:}\\
\multicolumn{2}{l}{{\small $z$ band: a1 =   0.742 , a2 = --0.751 , a3 = 1.200, a4 = --0.498}}\\ \hline
\end{tabular} 
\end{table}

\section{WASP-132}
WASP-132 is a $V$ = 12.4, K4 star with a metallicity of [Fe/H] = +0.22 
$\pm$ 0.13.  The transit $\log g_{*}$ of 4.61 $\pm$ 0.02 is consistent with the spectroscopic  $\log g_{*}$ of 4.6 $\pm$ 0.1.  The evolutionary comparison (Fig.~\ref{AgeMassFig}) gives an age of $>$\,0.9 Gyr.   

The radial-velocities show excess scatter, which may be due to magnetic activity. There is also a suggestion in Fig.~5 of a possible  correlation of the bisector with orbital phase. This is partly due to a possible longer-term trend to both lower radial velocities and bisectors over the span of the observations. The radial velocities decrease by 60 m s$^{-1}$ over time, though this is unreliable owing to the CORALIE upgrade midway through the dataset. If we analyse the data before and after the upgrade separately we find no significant correlation between the bisector and the radial-velocity value. 

A possible rotational modulation with a period of 33 $\pm$ 3 d, and an amplitude of 0.4-mmag, is seen in 3 out of 5 seasons of WASP data (Fig.~6), while a possible modulation at half this period is seen in a 4th dataset.  This is close to the limit detectable with WASP data (for the other stars we're quoting upper limits in the range 0.5--1.5 mmag), and so is not fully reliable.   

A rotational period of  33 $\pm$ 3 d would indicate a gyrochronological age of  2.2 $\pm$ 0.3 Gyr \citep{2007ApJ...669.1167B}, which is consistent with the above evolutionary estimate. The period would also imply an equatorial velocity of 1.1 $\pm$ 0.2 \kmps, which is consistent with the observed (but poorly constrained) \vsini\ value of 0.9 $\pm$ 0.8 \kmps. 

The planet, WASP-132b, has a low-mass and a modest radius compared to many hot Jupiters (0.41 M$_{\rm Jup}$; 0.87 R$_{\rm Jup}$).  With an orbital period of  7.1 d around a K4 star it is among the least irradiated of the WASP planets. The equilibrium temperature is estimated at only 763 $\pm$ 16 K.  Of WASP systems, only WASP-59b \citep{2013A&A...549A.134H}, in a 7.9-d orbit around a K5V, has a lower temperature of 670 $\pm$ 35 K. HATS-6b \citep{2015AJ....149..166H}, in a 3.3-d orbit around an M1V star, is also cooler (713 $\pm$ 5 K), but all other cooler gas giants have orbital periods of greater than 10 d.

\clearpage

\begin{figure}
\hspace*{2mm}\includegraphics[width=8.5cm]{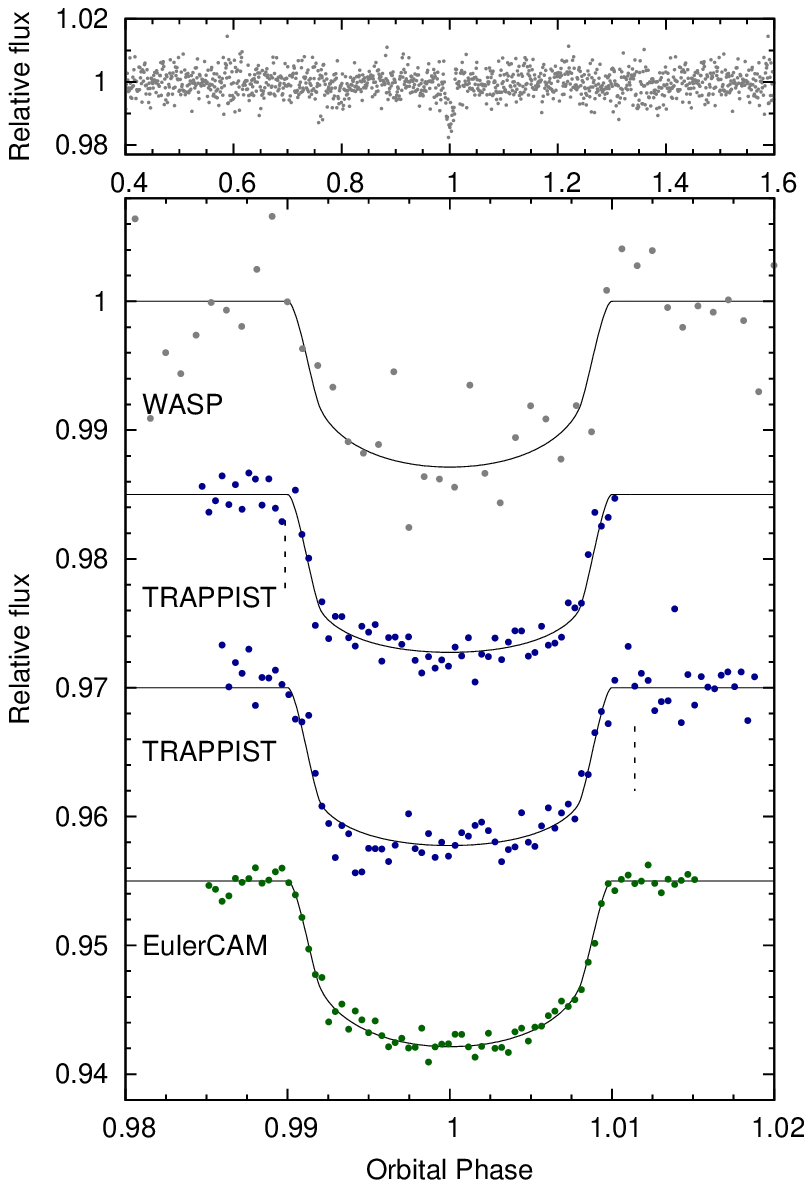}\\ [-2mm]
\caption{WASP-139b discovery photometry, as for Fig.~1.}
\end{figure}

\begin{figure}
\hspace*{2mm}\includegraphics[width=8.5cm]{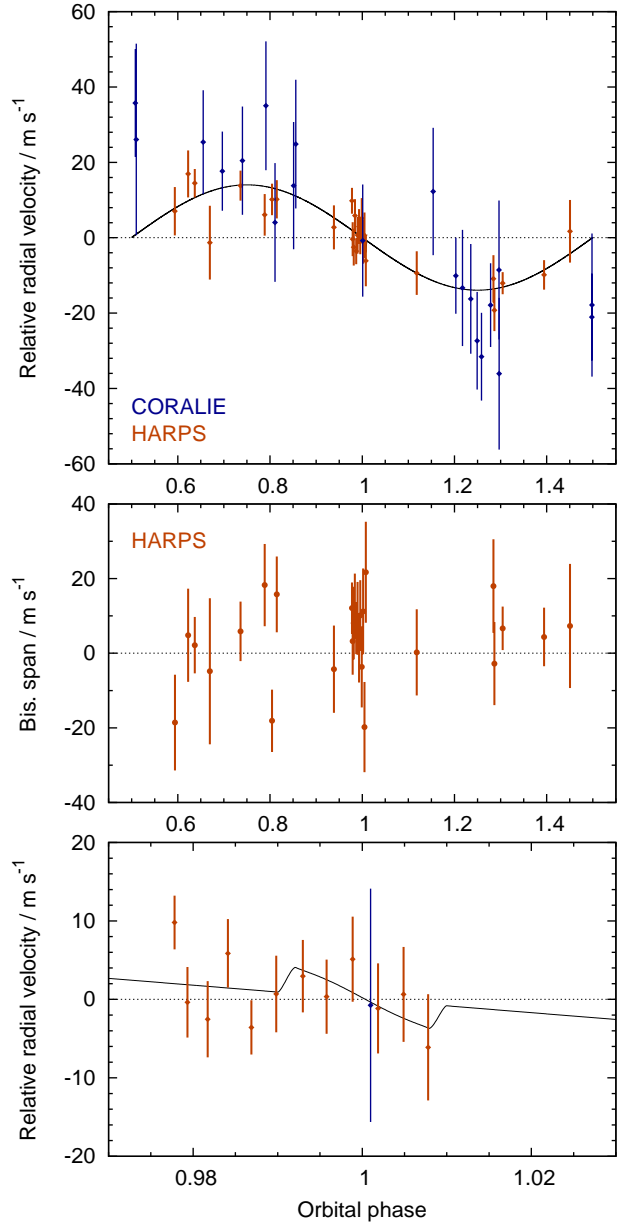}\\ [-2mm]
\caption{WASP-139b radial velocities and bisector spans, as for Fig.~2. We have not plotted 3 RV points with very large errors. We also show only the HARPS bisectors, which have smaller errors.  The lowest panel is a larger-scale view of the transit and the R--M effect.}
\end{figure}

\begin{table}
\caption{System parameters for WASP-139.}  
\begin{tabular}{lc}
\multicolumn{2}{l}{1SWASP\,J031814.91--411807.4}\\
\multicolumn{2}{l}{2MASS\,03181493--4118077}\\
\multicolumn{2}{l}{RA\,=\,03$^{\rm h}$18$^{\rm m}$14.93$^{\rm s}$, 
Dec\,=\,--41$^{\circ}$18$^{'}$07.7$^{''}$ (J2000)}\\
\multicolumn{2}{l}{$V$ mag = 12.4}  \\ 
\multicolumn{2}{l}{Rotational modulation\ \ \ $<$\,1 mmag (95\%)}\\
\multicolumn{2}{l}{pm (RA) --16.7\,$\pm$\,1.3 (Dec) 23.6\,$\pm$\,3.3 mas/yr}\\
\hline
\multicolumn{2}{l}{Stellar parameters from spectroscopic analysis.\rule[-1.5mm]{
0mm}{2mm}} \\ \hline 
Spectral type & K0 \\
$T_{\rm eff}$ (K)  & 5300  $\pm$ 100  \\
$\log g$      & 4.5 $\pm$ 0.1    \\
$v\,\sin i$ (km\,s$^{-1}$)     &    4.2 $\pm$ 1.1     \\
{[Fe/H]}   &   +0.20 $\pm$ 0.09     \\
log A(Li)  &    $<$ 0.5   \\
Age (Lithium) [Gy]  &  $\ga$ 0.5          \\
Age (Gyro) [Gy]     & $<\,0.5^{+0.4}_{-0.3}$   \\ 
Distance [pc]  & 230 $\pm$ 40   \\ \hline 
\multicolumn{2}{l}{Parameters from MCMC analysis.\rule[-1.5mm]{0mm}{3mm}} \\
\hline 
$P$ (d) & 5.924262  $\pm$ 0.000004 \\
$T_{\rm c}$ (HJD)\,(UTC) & 2457196.7933 $\pm$ 0.0003 \\
$T_{\rm 14}$ (d) & 0.118 $\pm$ 0.001 \\
$T_{\rm 12}=T_{\rm 34}$ (d) & 0.012 $\pm$ 0.002 \\
$\Delta F=R_{\rm P}^{2}$/R$_{*}^{2}$ & 0.0107 $\pm$ 0.0003 \\
$b$ & 0.33 $\pm$ 0.14  \\
$i$ ($^\circ$)  & 88.9 $\pm$ 0.5 \\
$K_{\rm 1}$ (km s$^{-1}$) & 0.0140 $\pm$ 0.0014 \\
$\gamma$ (km s$^{-1}$)  & --12.996 $\pm$ 0.001 \\
$e$ & 0 (adopted) ($<$\,0.28 at 2$\sigma$) \\ 
$M_{\rm *}$ (M$_{\rm \odot}$) & 0.92 $\pm$ 0.10 \\
$R_{\rm *}$ (R$_{\rm \odot}$) & 0.80 $\pm$ 0.04 \\
$\log g_{*}$ (cgs) & 4.59 $\pm$ 0.06 \\
$\rho_{\rm *}$ ($\rho_{\rm \odot}$) & 1.8 $\pm$ 0.2 \\
$T_{\rm eff}$ (K) & 5310 $\pm$ 90 \\
$M_{\rm P}$ (M$_{\rm Jup}$) & 0.117 $\pm$ 0.017 \\
$R_{\rm P}$ (R$_{\rm Jup}$) & 0.80 $\pm$ 0.05 \\
$\log g_{\rm P}$ (cgs) & 2.62 $\pm$ 0.06 \\
$\rho_{\rm P}$ ($\rho_{\rm J}$) & 0.23 $\pm$ 0.04 \\
$a$ (AU)  & 0.062 $\pm$ 0.002 \\
$T_{\rm P, A=0}$ (K) & 910 $\pm$ 30 \\ [0.5mm] \hline 
\multicolumn{2}{l}{Errors are 1$\sigma$; Limb-darkening coefficients were:}\\
\multicolumn{2}{l}{{\small $r$ band: a1 = 0.712, a2 = --0.642, a3 = 1.321, 
a4 = --0.598}}\\ 
\multicolumn{2}{l}{{\small $z$ band: a1 =   0.721 , a2 = --0.671 , a3 = 1.104, a4 = --0.494}}\\ \hline
\end{tabular} 
\end{table}

\section{WASP-139}
WASP-139 is a $V$ = 12.4, K0 star with a metallicity of [Fe/H] = +0.20 $\pm$ 0.09. The transit $\log g_{*}$ of 4.59 $\pm$ 0.06 is consistent with the spectroscopic  $\log g_{*}$ of 4.5 $\pm$ 0.1.    The gyrochonological age constraint and the lack of lithium imply a relatively young star of $\sim$\,0.5 Gyr. 

The stellar density resulting from the transit analysis (1.8 $\pm$ 0.2\,$\rho_{\odot}$; 0.92 M$_{\rm \odot}$, 0.80 R$_{\rm \odot}$)  puts the star below the main sequence and is only marginally consistent with the evolutionary models of \citet{2015A&A...575A..36M}.  The same has been found for HAT-P-11 \citep{2010ApJ...710.1724B} and possibly also for WASP-89 \citep{2015AJ....150...18H}. For a discussion of this see \citet{2015A&A...577A..90M}, who suggested that such stars might be helium-rich.  

The planet, WASP-139b, has a mass of only 0.12 $\pm$ 0.02 M$_{\rm Jup}$, making it the lowest-mass WASP discovery yet. With a radius of 0.80 R$_{\rm Jup}$, and thus a low density of 0.23 $\pm$ 0.04 $\rho_{\rm Jup}$, the large scale height makes  WASP-139b a good target for atmospheric characterisation.

Owing to the small planet mass, and thus the low reflex velocity, we obtained HARPS data in order to better parametrise the system. This included observations of the Rossiter--McLaughlin effect through transit (Fig.~8). If the orbit were aligned, and taking values for the \vsini\ and impact parameter from Table~5, we'd expect an R--M effect of order 30 m s$^{-1}$ (e.g.~\citealt{2007ApJ...655..550G}). The HARPS data indicate a much lower value, though owing to the relatively large errors the fit is effectively unconstrained and thus we do not report parameters such as the alignment angle. 

WASP-139b is most similar to two recent discoveries by the HATSouth
project, HATS-7b \citep{2015ApJ...813..111B} and HATS-8b
\citep{2015AJ....150...49B}.  HATS-7b is a 0.12 M$_{\rm Jup}$ planet
with a radius of 0.56 R$_{\rm Jup}$ in a 3.2-d orbit.  The host stars
are also similar (HATS-7 is a V = 13.3, K dwarf, T$_{\rm eff}$ = 4985,
[Fe/H] = +0.25; HATS-8 is a V = 14.0 G dwarf, T$_{\rm eff}$ = 5680,
[Fe/H] = +0.21; whereas WASP-139 is a V = 12.4 K0 dwarf, T$_{\rm eff}$
= 5300, [Fe/H] = +0.20).  The HATS project have called such systems
``super-Neptunes'', and, as now the brightest example, the WASP-139
system will be important for studying such objects.


\begin{figure}
\hspace*{2mm}\includegraphics[width=8.5cm]{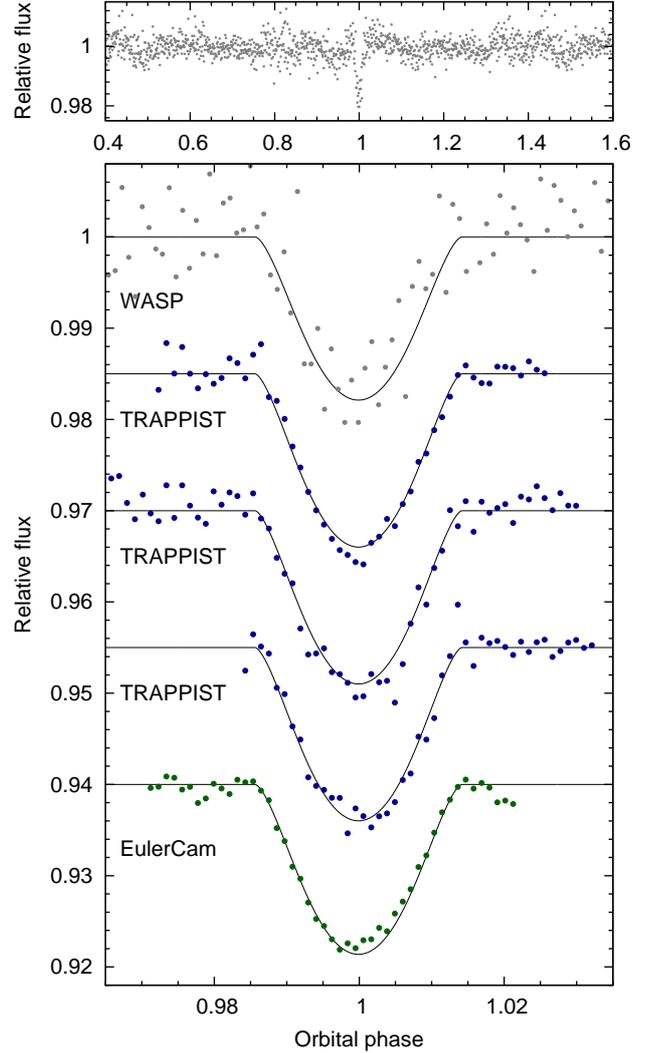}\\ [-2mm]
\caption{WASP-140b photometry, as for Fig.~1.}
\end{figure}

\begin{figure}
\hspace*{2mm}\includegraphics[width=8.5cm]{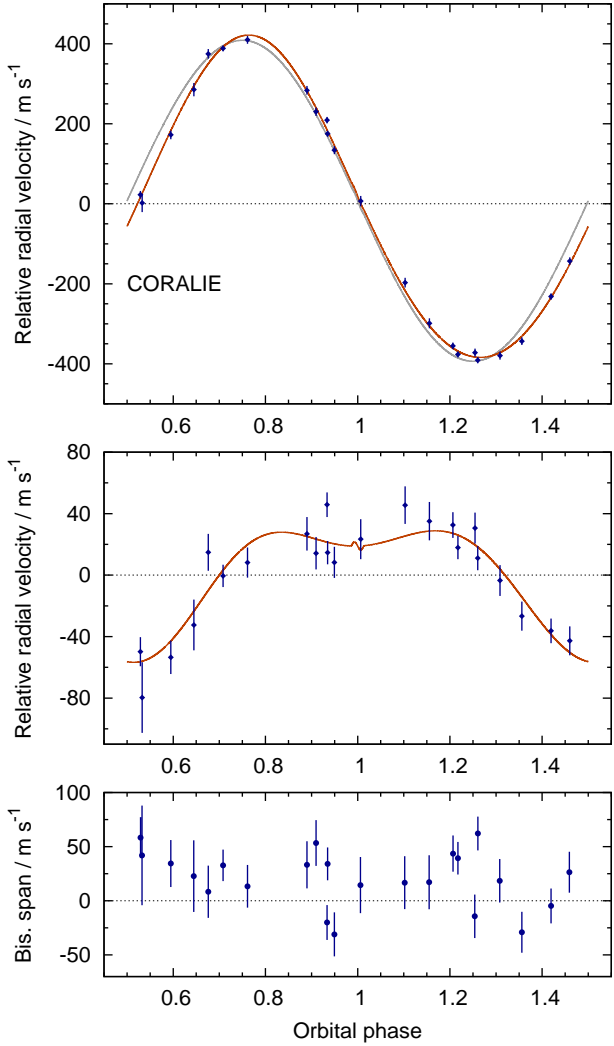}\\ [-2mm]
\caption{WASP-140b radial velocities, as for Figs.~2. (Top) The dark-orange radial velocity curve is the best-fitting eccentric orbit; the grey line is the best circular orbit for comparison. (Middle) The RVs after subtracting the circular orbit. (Lowest) The bisector spans.}
\end{figure}

\begin{figure}
\hspace*{0mm}\includegraphics[width=9cm]{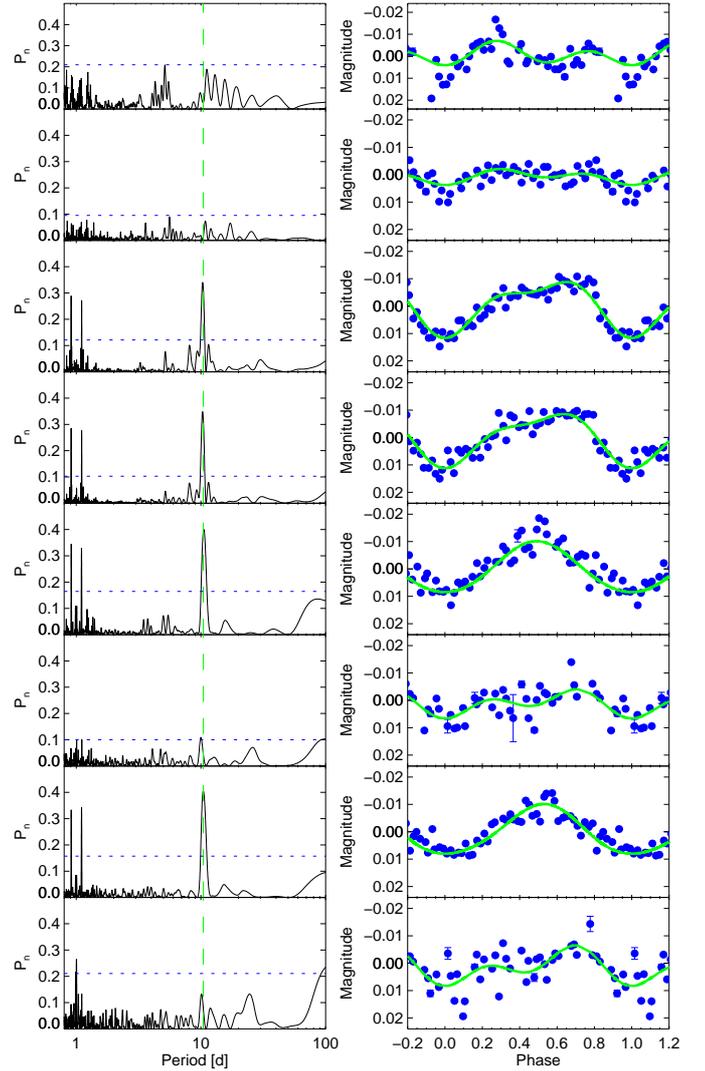}\\ [-2mm]
\caption{Rotational modulation in the WASP data for WASP-140.  The left-hand panels show peridograms for independent data sets (a season of data in a given camera over 2006 to 2011). The blue dotted line is a false-alarm probability of 0.001.  The right-hand panels show the data for each season folded on the 10.4-d period; the green line is a harmonic-series fit to the data}
\end{figure}

\begin{table}
\caption{System parameters for WASP-140.}  
\begin{tabular}{lc}
\multicolumn{2}{l}{1SWASP\,J040132.53--202703.9}\\
\multicolumn{2}{l}{2MASS\,04013254--2027039}\\
\multicolumn{2}{l}{RA\,=\,04$^{\rm h}$01$^{\rm m}$32.54$^{\rm s}$, 
Dec\,=\,--20$^{\circ}$27$^{'}$03.9$^{''}$ (J2000)}\\
\multicolumn{2}{l}{$V$ mag = 11.1}  \\ 
\multicolumn{2}{l}{Rotational modulation: 5--9 mmag at 10.4\,$\pm$\,0.1 d}\\
\multicolumn{2}{l}{pm (RA) --23.2\,$\pm$\,1.7 (Dec) 17.8\,$\pm$\,1.1 mas/yr}\\
\hline
\multicolumn{2}{l}{Stellar parameters from spectroscopic analysis.\rule[-1.5mm]{
0mm}{2mm}} \\ \hline 
Spectral type & K0 \\
$T_{\rm eff}$ (K)  & 5300  $\pm$ 100  \\
$\log g$      & 4.2 $\pm$ 0.1    \\
$v\,\sin i$ (km\,s$^{-1}$)     &    3.1 $\pm$ 0.8     \\
{[Fe/H]}   &   +0.12 $\pm$ 0.10     \\
log A(Li)  &    $<$ 0.4   \\
Age (Lithium) [Gy]  &   $\ga$ 0.5        \\
Age (Gyro) [Gy]     &  $<\,1.6^{+1.4}_{-0.9}$  \\ 
Distance [pc]  & 180 $\pm$ 30   \\ \hline 
\multicolumn{2}{l}{Parameters from MCMC analysis.\rule[-1.5mm]{0mm}{3mm}} \\
\hline 
$P$ (d) & 2.2359835 $\pm$ 0.0000008 \\
$T_{\rm c}$ (HJD) & 2456912.35105 $\pm$ 0.00015 \\
$T_{\rm 14}$ (d) & 0.0631 $\pm$ 0.0009 \\
$T_{\rm 12}=T_{\rm 34}$ (d) & (undefined) \\
$\Delta F=R_{\rm P}^{2}$/R$_{*}^{2}$ & 0.0205 $^{+0.002}_{-0.0005}$  \\
$b$ & 0.93 $^{+0.07}_{-0.03}$ \\
$i$ ($^\circ$)  & 83.3 $^{+0.5}_{-0.8}$ \\
$K_{\rm 1}$ (km s$^{-1}$) & 0.403 $\pm$ 0.003 \\
$\gamma$ (km s$^{-1}$)  & 2.125 $\pm$ 0.003 \\
$e\cos\omega$ & 0.0468 $\pm$ 0.0035 \\
$e\sin\omega$ & --0.003 $\pm$ 0.006 \\
$e$ & 0.0470 $\pm$ 0.0035 \\
$\omega$ ($^\circ$) & --4 $\pm$ 8 \\
$\phi_{\rm mid-occultation}$ & 0.530 $\pm$ 0.002 \\
$M_{\rm *}$ (M$_{\rm \odot}$) & 0.90 $\pm$ 0.04 \\
$R_{\rm *}$ (R$_{\rm \odot}$) & 0.87 $\pm$ 0.04 \\
$\log g_{*}$ (cgs) & 4.51 $\pm$ 0.04 \\
$\rho_{\rm *}$ ($\rho_{\rm \odot}$) & 1.38 $\pm$ 0.18 \\
$T_{\rm eff}$ (K) & 5260 $\pm$ 100 \\
$M_{\rm P}$ (M$_{\rm Jup}$) & 2.44 $\pm$ 0.07 \\
$R_{\rm P}$ (R$_{\rm Jup}$) & 1.44 $^{+0.42}_{-0.18}$\\
$\log g_{\rm P}$ (cgs) & 3.4 $\pm$ 0.2 \\
$\rho_{\rm P}$ ($\rho_{\rm J}$) & 0.8 $\pm$ 0.4 \\
$a$ (AU)  & 0.0323 $\pm$ 0.0005 \\
$T_{\rm P, A=0}$ (K) & 1320 $\pm$ 40 \\ [0.5mm] \hline 
\multicolumn{2}{l}{Errors are 1$\sigma$; Limb-darkening coefficients were:}\\
\multicolumn{2}{l}{{\small $z$ band: a1 = 0.725, a2 = --0.684, a3 = 1.121, 
a4 = --0.496}}\\ 
\multicolumn{2}{l}{{\small $I$ band: a1 = 0.786, a2 = --0.811 , a3 = 1.320, a4 = --0.573}}\\ \hline
\end{tabular} 
\end{table}

\section{WASP-140}
WASP-140A is a $V$ = 11.1, K0 star with a metallicity of [Fe/H] = +0.12 
$\pm$ 0.10.  The transit $\log g_{*}$ of 4.51 $\pm$ 0.04 is higher than the spectroscopic  $\log g_{*}$ of 4.2 $\pm$ 0.1. In such cases we regard the transit value as the more reliable, given the systematic uncertainties in $\log g_{*}$ estimates in such spectra (e.g.~\citealt{2015ApJ...805..126B} report discrepancies as big as 0.3 dex).    

A second star, WASP-140B, is fainter by 2.01 $\pm$ 0.02 magnitudes and is 7.24 $\pm$ 0.01 arcsecs from WASP-140 at a position angle of 77.4 $\pm$ 0.1 degrees (values from the EulerCAM observation on 2015-09-01 with a $I_{c}$ filter).  The TRAPPIST and EulerCAM transit photometry used a small aperture that excluded this star.   The 2MASS colours of WASP-140B ($J$ =  11.09 $\pm$ 0.03; $H$ = 10.46 $\pm$ 0.02; $Ks$ = 10.27 $\pm$ 0.03) are consistent with it being physically associated with WASP-140A ($J$ =  9.61 $\pm$ 0.03; $H$ = 9.24 $\pm$ 0.02; $Ks$ = 9.17 $\pm$ 0.03), and so it is possible that the two stars form a binary. There are no proper motion values listed for WASP-140B in UCAC4.

The WASP data on WASP-140 show a clear rotational modulation with a period of 10.4 $\pm$ 0.1 days and an  amplitude varying between 5 and 9 mmag (Fig.~11), implying that it is magnetically active.  The WASP aperture includes both stars, so it is not certain which star is the variable, though if it were WASP-140B then the amplitude would have to be 6 times higher, which is less likely.  There is also evidence of a star spot in each of the two lowest transit lightcurves in Fig.~9, which would imply that WASP-140A is magentically active.    

The 10.4-d rotational period would imply a young gyrochronological age for WASP-140A of 0.42 $\pm$ 0.06 Gyr \citep{2007ApJ...669.1167B}.   This is inconsistent with the evolutionary comparison (Fig.~\ref{AgeMassFig}), which suggests a likeliest age of 8 Gyr with a lower bound of 1.7 Gyr.   This inconsistency suggests that WASP-140A has been spun up by the presence of the massive, closely orbiting planet (see the discussion in \citealt{2014MNRAS.442.1844B}). 

The rotational period equates to an equatorial velocity of 6.3 $\pm$ 0.9 \kmps. Comparing this to the observed \vsini\ value of 3.1 $\pm$ 0.8 \kmps\ suggests a misaligned system, with the star's spin axis at an inclination of 30$^{\circ}$ $\pm$ 15$^{\circ}$.

The planet WASP-140Ab has a mass of 2.4 M$_{\rm Jup}$ and is in a 2.2-day orbit. The transit is grazing, with an impact parameter of  0.93 $^{+0.07}_{-0.03}$. Other WASP planets that are grazing are WASP-67b (\citealt{2012MNRAS.426..739H}; \citealt{2014A&A...568A.127M}) and WASP-34b \citep{2011A&A...526A.130S}.  Since it is possible that not all of the planet is transiting the star its radius is ill-constrained at 1.44 $^{+0.42}_{-0.18}$ R$_{\rm Jup}$.   

\subsection{WASP-140Ab's eccentric orbit}
The orbit of WASP-140Ab is eccentric with $e = 0.0470 \pm 0.0035$.  A Lucy--Sweeney test shows this to be significantly non-zero with $>\,99.9$\%\ confidence.   Being significantly eccentric at an orbital period as short as 2.2 days is unusual in a hot Jupiter. For comparison,  WASP-14b \citep{2009MNRAS.392.1532J} also has an eccentric 2.2-day orbital period, but is a much more massive planet at 7.7 M$_{\rm Jup}$.  WASP-89b \citep{2015AJ....150...18H} has an eccentric 3.4-d orbit and is also more massive at 5.9 M$_{\rm Jup}$.

The circularisation timescale for a hot Jupiter can be estimated from (\citealt{2006ApJ...649.1004A}, eqn 3):

\begin{eqnarray*}
\tau_{\rm cir}\ \approx &\ 1.6~{\rm Gyr} \times \left(\frac{Q_{\rm P}}{10^6}\right) \times \left(\frac{M_{\rm P}}{M_{\rm Jup}}\right) \times \left(\frac{M_*}{\Msolar}\right)^{-3/2} \nonumber \\  & \times \left(\frac{R_{\rm P}}{R_{\rm Jup}}\right)^{-5} \times \left(\frac{a}{0.05~{\rm AU}}\right)^{13/2}
\end{eqnarray*} 

Using a value of the quality factor, ${Q_{\rm P}}$, of 10$^{5}$ (e.g.\ \citealt{2012arXiv1209.5724S}), and the parameters of Table~6, gives a circularisation timescale of $\sim$ 5 Myr.    Note, however, the strong dependence on $R_{\rm P}$, which is poorly contrained in WASP-140Ab owing to the grazing transit.    Pushing ${Q_{\rm P}}$ up to  10$^{6}$, and taking the parameters at their 1-sigma boundaries to lengthen the timescale allows values of $\sim$ 100 Myr.  This is still short compared to the likely age of the host star, and suggests that WASP-140b has only relatively recently arrived in its current orbit.  

Comparing to other hot-Jupiters, using the above  equation and parameters tabulated in TEPCat \citep{2011MNRAS.417.2166S},  we find that WASP-140Ab has the shortest circularisation timescale of all hot Jupiters that are in clearly eccentric orbits (where we adopt a 3$\sigma$ threshold). Using the best-fit parameters of Table~6 for WASP-140Ab, and adopting  ${Q_{\rm P}}$ = 10$^{5}$,  gives $\log(\tau_{\rm cir})$ = 6.6. 

A timescale of $\log(\tau_{\rm cir})$ = 6.1 is obtained for WASP-18b \citep{2009Natur.460.1098H}, which has been reported as having a small but significant eccentricity of $e$ = 0.008 $\pm$ 0.001 \citep{2010A&A...524A..25T}. However, this apparent eccentricity might instead be an effect of the tidal bulge on WASP-18, which is the biggest of any known hot-Jupiter system (see \citealt{2012MNRAS.422.1761A}).

The next shortest timescale is $\log(\tau_{\rm cir})$ = 6.8 for HAT-P-13b \citep{2009ApJ...707..446B}. From {\it Spitzer\/} observations of the planetary occulation, \citet{2016ApJ...821...26B} report a significant eccentricity of $e = 0.007 \pm 0.001$.  In this system, however, the eccentricity of the hot Jupiter HAT-P-13b is likely being maintained by the perturbative effect of  HAT-P-13c, a 14 M$_{\rm Jup}$ outer planet in a highly eccentric ($e = 0.66$) 446-day orbit \citep{2010ApJ...718..575W}.

The smallest timescale for any other hot Jupiter that is indisputably
eccentric is likely that for WASP-14b at $\log(\tau_{\rm cir})$ =
7.6. This is an order of magnitude longer than that for WASP-140Ab,
which implies that WASP-140Ab is unusual.  Tidal heating has long been
proposed as a possible cause of the inflated radii of many hot
Jupiters (e.g.~\citealt{2013arXiv1304.4121S} and references therein),
and may help to explain the fact that WASP-140Ab has a bloated radius
despite being relatively massive.  It will be worth obtaining better
transit photometry of WASP-140, in order to better constrain the
parameters, and also worth looking for an outer planet that might be
maintaining the eccentricity.

It's also worth noting that short-period, massive and eccentric
planets are rare around K stars. WASP-89b is the previously known
example, a 6 M$_{\rm Jup}$ planet in a 3.36-d orbit with an
eccentricity of 0.192 $\pm$ 0.009 around a K3 star
\citep{2015AJ....150...18H}. The magnetic activity of both stars,
WASP-89 and WASP-140A, might be related to the presence of the
eccentric, short-period planet (e.g.~\citealt{2014A&A...565L...1P}).

\begin{figure}
\hspace*{2mm}\includegraphics[width=8.5cm]{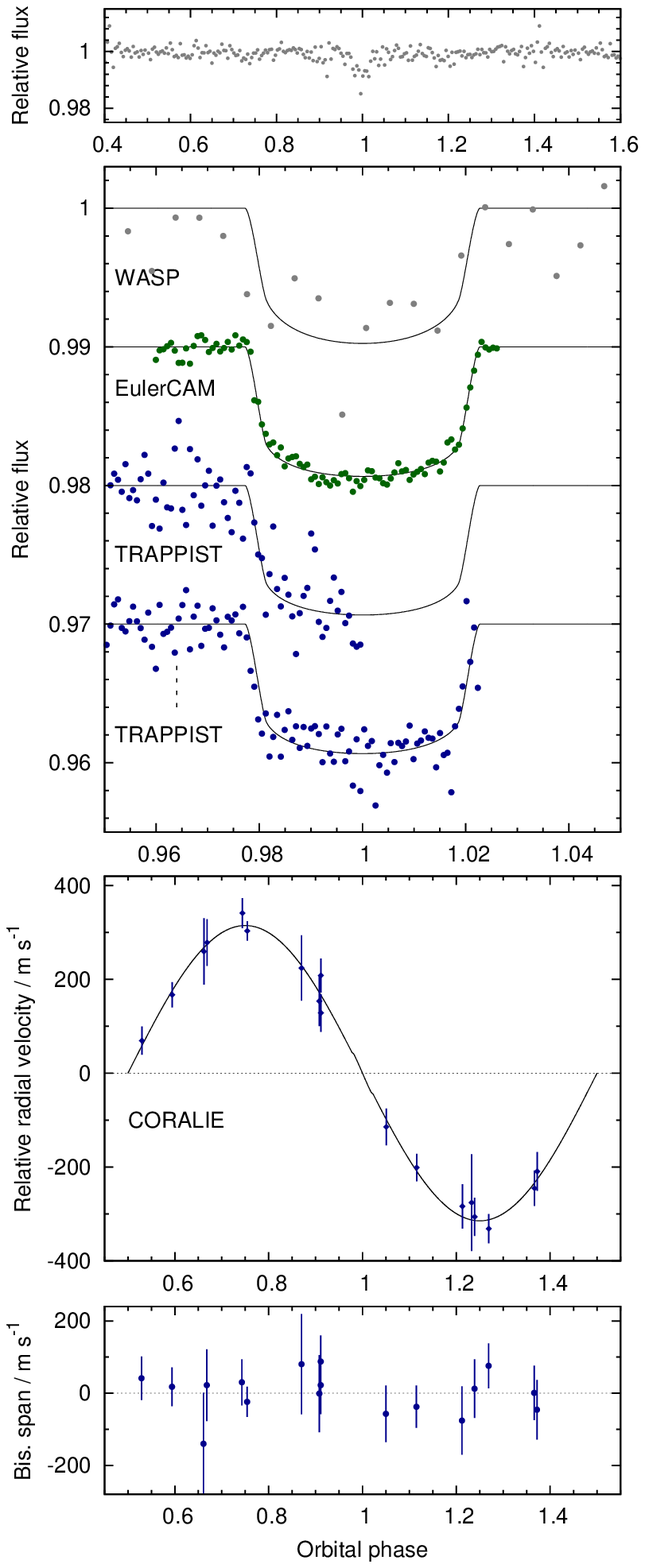}\\ [-2mm]
\caption{WASP-141b discovery data, as for Figs.~1 \&\ 2.}
\end{figure}

\begin{table}
\caption{System parameters for WASP-141.}  
\begin{tabular}{lc}
\multicolumn{2}{l}{1SWASP\,J044717.84--170654.6}\\
\multicolumn{2}{l}{2MASS\,04471785--1706545}\\
\multicolumn{2}{l}{RA\,=\,04$^{\rm h}$47$^{\rm m}$17.85$^{\rm s}$, 
Dec\,=\,--17$^{\circ}$06$^{'}$54.5$^{''}$ (J2000)}\\
\multicolumn{2}{l}{$V$ mag = 12.4}  \\ 
\multicolumn{2}{l}{Rotational modulation\ \ \ $<$\,1.5 mmag (95\%)}\\
\multicolumn{2}{l}{pm (RA) 4.2\,$\pm$\,0.9 (Dec) --3.1\,$\pm$\,2.6 mas/yr}\\
\hline
\multicolumn{2}{l}{Stellar parameters from spectroscopic analysis.\rule[-1.5mm]{
0mm}{2mm}} \\ \hline 
Spectral type & F9 \\
$T_{\rm eff}$ (K)  & 6050  $\pm$ 120  \\
$\log g$      & 4.20 $\pm$ 0.15    \\
$v\,\sin i$ (km\,s$^{-1}$)     &    3.9 $\pm$ 0.8     \\
{[Fe/H]}   &   +0.29 $\pm$ 0.09     \\
log A(Li)  &    1.75 $\pm$ 0.12   \\
Age (Lithium) [Gy]  &   $\ga$ 5           \\
Age (Gyro) [Gy]     &  $<\,4.0^{+4.6}_{-2.4}$   \\ 
Distance [pc]  & 570 $\pm$ 150  \\ \hline 
\multicolumn{2}{l}{Parameters from MCMC analysis.\rule[-1.5mm]{0mm}{3mm}} \\
\hline 
$P$ (d) & 3.310651 $\pm$ 0.000005 \\
$T_{\rm c}$ (HJD)\,(UTC) & 2457019.5953 $\pm$ 0.0003 \\
$T_{\rm 14}$ (d) & 0.150 $\pm$ 0.001 \\
$T_{\rm 12}=T_{\rm 34}$ (d) & 0.014 $\pm$ 0.001 \\
$\Delta F=R_{\rm P}^{2}$/R$_{*}^{2}$ & 0.0083 $\pm$ 0.0002 \\
$b$ & 0.31 $\pm$ 0.12 \\
$i$ ($^\circ$) & 87.6 $\pm$ 1.3 \\
$K_{\rm 1}$ (km s$^{-1}$) & 0.315 $\pm$ 0.015 \\
$\gamma$ (km s$^{-1}$)  & 33.828 $\pm$ 0.009 \\
$e$ & 0 (adopted) ($<$\,0.06 at 2$\sigma$) \\ 
$M_{\rm *}$ (M$_{\rm \odot}$) & 1.25 $\pm$ 0.06 \\
$R_{\rm *}$ (R$_{\rm \odot}$) & 1.37 $\pm$ 0.07 \\
$\log g_{*}$ (cgs) & 4.26 $\pm$ 0.04 \\
$\rho_{\rm *}$ ($\rho_{\rm \odot}$) & 0.49 $\pm$ 0.07 \\
$T_{\rm eff}$ (K) & 5900 $\pm$ 120 \\
$M_{\rm P}$ (M$_{\rm Jup}$) & 2.69 $\pm$ 0.15 \\
$R_{\rm P}$ (R$_{\rm Jup}$) & 1.21 $\pm$ 0.08 \\
$\log g_{\rm P}$ (cgs) & 3.62 $\pm$ 0.05 \\
$\rho_{\rm P}$ ($\rho_{\rm J}$) & 1.49 $\pm$ 0.25 \\
$a$ (AU)  & 0.0469 $\pm$ 0.0007 \\
$T_{\rm P, A=0}$ (K) & 1540 $\pm$ 50 \\ [0.5mm] \hline 
\multicolumn{2}{l}{Errors are 1$\sigma$; Limb-darkening coefficients were:}\\
\multicolumn{2}{l}{{\small $z$ band: a1 = 0.616, a2 = --0.305, a3 = 0.635, 
a4 = --0.331}}\\ \hline
\end{tabular} 
\end{table}

\begin{figure}
\hspace*{2mm}\includegraphics[width=8.5cm]{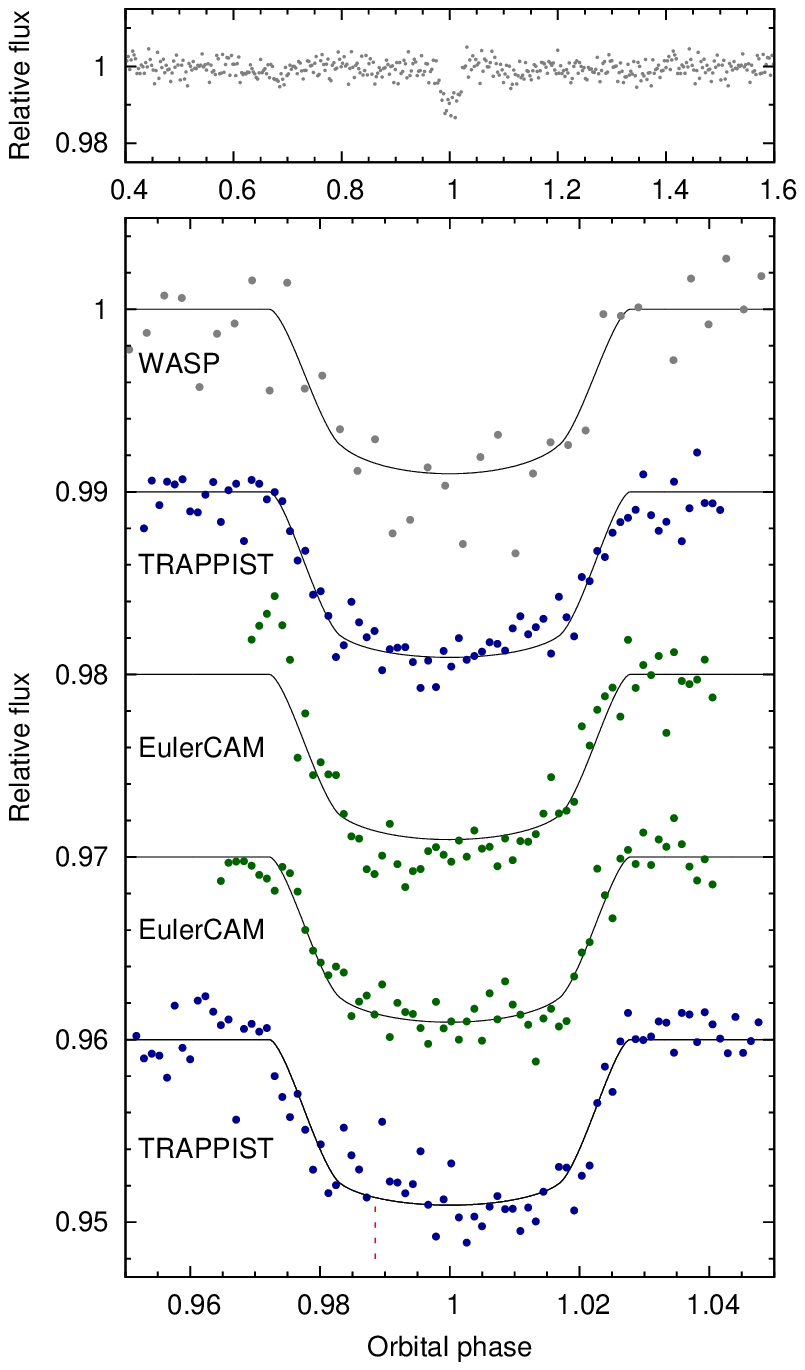}\\ [-2mm]
\caption{WASP-142b discovery photometry, as for Fig.~1.}
\end{figure}

\section{WASP-141}
WASP-141 is a $V$ = 12.4, F9  star with a metallicity of [Fe/H] =  +0.29 $\pm$ 0.09.  The transit $\log g_{*}$ of 4.26 $\pm$ 0.06 is consistent with the spectroscopic value of 4.20 $\pm$ 0.15.   The evolutionary comparison (Fig.~\ref{AgeMassFig}) gives an age estimate of 1.5--5.6 Gyr. This is compatible with the gyrochronological estimate of $<\,4.0^{+4.6}_{-2.4}$ Gyr and marginally consistent with the lithium age of $\ga$ 5 Gyr.  

The planet, WASP-141b is a 2.7 M$_{\rm Jup}$, 1.2 R$_{\rm Jup}$ planet in a 3.3-d orbit. WASP-141 appears to be a typical hot-Jupiter system.

\begin{table}
\caption{System parameters for WASP-142.}  
\begin{tabular}{lc}
\multicolumn{2}{l}{1SWASP\,J092201.43--235645.8}\\
\multicolumn{2}{l}{2MASS\,09220153--2356462}\\
\multicolumn{2}{l}{RA\,=\,09$^{\rm h}$22$^{\rm m}$01.53$^{\rm s}$, 
Dec\,=\,--23$^{\circ}$56$^{'}$46.2$^{''}$ (J2000)}\\
\multicolumn{2}{l}{$V$ mag = 12.3}  \\ 
\multicolumn{2}{l}{Rotational modulation\ \ \ $<$\,1.5 mmag (95\%)}\\
\multicolumn{2}{l}{pm (RA) --3.1\,$\pm$\,3.5 (Dec) 3.7\,$\pm$\,3.1 mas/yr}\\
\hline
\multicolumn{2}{l}{Stellar parameters from spectroscopic analysis.\rule[-1.5mm]{
0mm}{2mm}} \\ \hline 
Spectral type & F8 \\
$T_{\rm eff}$ (K)  & 6060  $\pm$ 150  \\
$\log g$      & 4.0 $\pm$ 0.2    \\
$v\,\sin i$ (km\,s$^{-1}$)     &    3.1 $\pm$ 1.4     \\
{[Fe/H]}   &   +0.26 $\pm$ 0.12     \\
log A(Li)  &    3.10 $\pm$ 0.09   \\
Age (Lithium) [Gy]  &   $\la$ 2          \\
Distance [pc]  &  840 $\pm$ 310   \\ \hline 
\multicolumn{2}{l}{Parameters from MCMC analysis.\rule[-1.5mm]{0mm}{3mm}} \\
\hline 
$P$ (d) & 2.052868 $\pm$ 0.000002 \\
$T_{\rm c}$ (HJD) & 2457007.7779 $\pm$ 0.0004 \\
$T_{\rm 14}$ (d) & 0.1117 $\pm$ 0.0016 \\
$T_{\rm 12}=T_{\rm 34}$ (d) & 0.022 $\pm$ 0.002 \\
$\Delta F=R_{\rm P}^{2}$/R$_{*}^{2}$ & 0.00916 $\pm$ 0.00026 \\
$b$ & 0.77 $\pm$ 0.02 \\
$i$ ($^\circ$)  & 80.2 $\pm$ 0.6 \\
$K_{\rm 1}$ (km s$^{-1}$) & 0.109 $\pm$ 0.010 \\
$\gamma$ (km s$^{-1}$)  & 47.126 $\pm$ 0.010 \\
$e$ & 0 (adopted) ($<$\,0.27 at 2$\sigma$) \\ 
$M_{\rm *}$ (M$_{\rm \odot}$) & 1.33 $\pm$ 0.08 \\
$R_{\rm *}$ (R$_{\rm \odot}$) & 1.64 $\pm$ 0.08 \\
$\log g_{*}$ (cgs) & 4.13 $\pm$ 0.04 \\
$\rho_{\rm *}$ ($\rho_{\rm \odot}$) & 0.30 $\pm$ 0.04 \\
$T_{\rm eff}$ (K) & 6010 $\pm$ 140 \\
$M_{\rm P}$ (M$_{\rm Jup}$) & 0.84 $\pm$ 0.09 \\
$R_{\rm P}$ (R$_{\rm Jup}$) & 1.53 $\pm$ 0.08 \\
$\log g_{\rm P}$ (cgs) & 2.91 $\pm$ 0.06 \\
$\rho_{\rm P}$ ($\rho_{\rm J}$) & 0.23 $\pm$ 0.05 \\
$a$ (AU)  & 0.0347 $\pm$ 0.0007 \\
$T_{\rm P, A=0}$ (K) & 2000 $\pm$ 60 \\ [0.5mm] \hline 
\multicolumn{2}{l}{Errors are 1$\sigma$; Limb-darkening coefficients were:}\\
\multicolumn{2}{l}{{\small $z$ band: a1 = 0.570, a2 = --0.130, a3 = 0.391, 
a4 = --0.233}}\\ 
\multicolumn{2}{l}{{\small $I$ band: a1 = 0.673 , a2 = --0.340 , a3 = 0.666, a4 = --0.340}}\\ \hline
\end{tabular} 
\end{table}

\section{WASP-142}
WASP-142A is a $V$ = 12.3, F8 star with a metallicity of [Fe/H] = +0.26 $\pm$ 0.12. The transit $\log g_{*}$ of 4.13 $\pm$ 0.04 is consistent with the spectroscopic value of 4.0 $\pm$ 0.2.  The evolutionary comparison (Fig.~\ref{AgeMassFig}) gives an age estimate of 2.2--7.0 Gyr.  The  lithium age is marginally inconsistent at $\la$ 2 Gyr. 

A second star, WASP-142B, is fainter by 1.86 $\pm$ 0.01 magnitudes and at 5.11 $\pm$ 0.01 arcsecs from WASP-142 at a position angle of $-$45.7 $\pm$ 0.1 degrees (values from an EulerCAM observation on 2014-12-13 with a $I_{c}$ filter).  The 2014 December EulerCAM transit photometry used an aperture including both stars, and we corrected the lightcurve for the dilution in the analysis.  The other EulerCAM transit and the two TRAPPIST transits used a smaller photometric aperture excluding the second star. 

 The 2MASS colours of WASP-142B ($J$ =  13.42 $\pm$ 0.04; $H$ = 13.03 $\pm$ 0.04; $Ks$ = 12.94 $\pm$ 0.03) are consistent with it being physically associated with WASP-142A ($J$ =  11.73 $\pm$ 0.03; $H$ = 11.48 $\pm$ 0.03; $Ks$ = 11.44 $\pm$ 0.03).     UCAC4, however, reports a very different proper motion for WASP-142B (pmRA = --99.1\,$\pm$\,2.1, pmDec = 98.3\,$\pm$\,2.2 mas/yr) than for WASP-142A (pmRA = --3.1\,$\pm$\,3.5, pmDec = 3.7\,$\pm$\,3.1 mas/yr), which, if reliable, would rule out a physical association.  

WASP-142Ab is a bloated planet of sub-Jupiter mass (1.53 R$_{\rm Jup}$; 0.84 M$_{\rm Jup}$) in a 2.1-d orbit.  Again, WASP-142 is a fairly typical hot-Jupiter system.

\clearpage

\begin{figure}
\hspace*{2mm}\includegraphics[width=8.5cm]{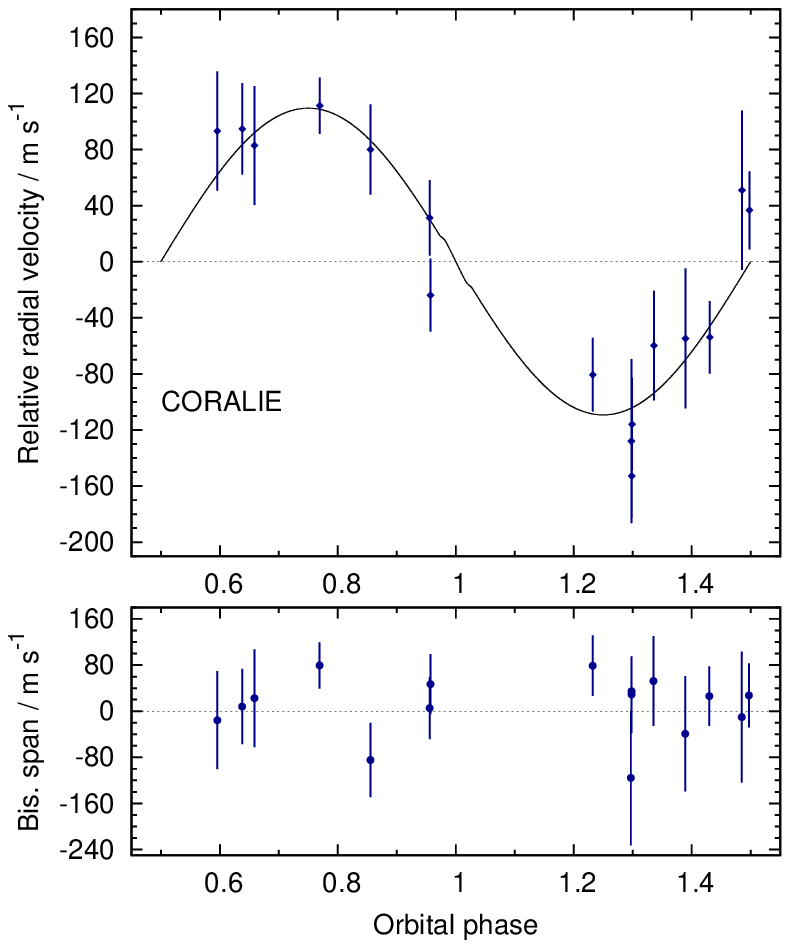}\\ [-2mm]
\caption{WASP-142b radial velocities and bisector spans, as for Fig.~2.}
\end{figure}

\begin{table}
 \caption{Bayesian mass and age estimates for the host stars.}
\label{AgeMassTable}
 \begin{tabular}{@{}lccc}
\hline
  \multicolumn{1}{@{}l}{Star} &
  \multicolumn{1}{c}{Likeliest age} &
  \multicolumn{1}{c}{95\%\ range} &
  \multicolumn{1}{c}{Mass} \\
  \multicolumn{1}{c}{} &
  \multicolumn{1}{c}{[Gyr]} &
  \multicolumn{1}{c}{[Gyr]} &
  \multicolumn{1}{c}{[\Msolar]} \\
\hline
 \noalign{\smallskip}
WASP-130 & 1.4 &  0.2--7.9 & 1.03 $\pm$ 0.04 \\
WASP-131 & 7.5 &  4.5--10.1 & 1.06 $\pm$ 0.06 \\
WASP-132 & 1.8 &  $>$ 0.9  & 0.80 $\pm$ 0.04 \\
WASP-139 & 0.0 &  $<$ 9.8 & 0.92 $\pm$ 0.04 \\
WASP-140 & 8.3 &  $>$ 1.7  & 0.90 $\pm$ 0.04 \\
WASP-141 & 3.2 &  1.5--5.6 & 1.25 $\pm$ 0.06 \\
WASP-142 & 3.6 &  2.2--7.0 & 1.32 $\pm$ 0.08 \\
\end{tabular}   
\end{table}

\begin{figure}
\mbox{\includegraphics[width=0.49\textwidth]{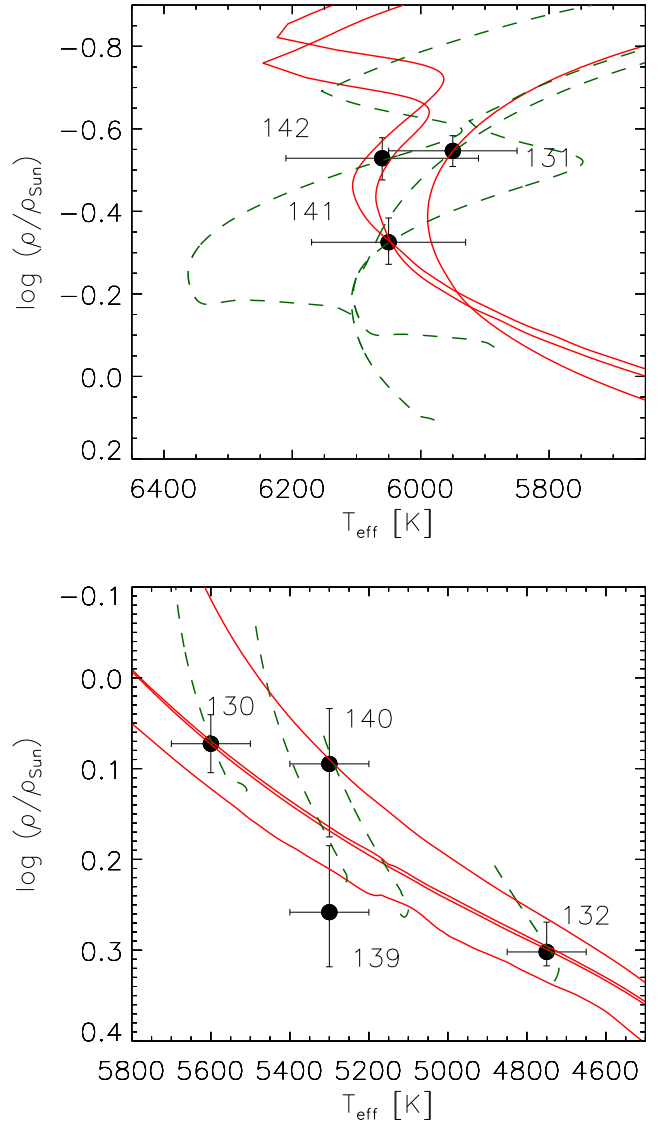}}
\caption{Mean stellar densities versus effective temperatures. Each planet is plotted with the mass track (green-dashed line) and isochrone (red line) for the best-fitting mass and age as listed in  Table~\ref{AgeMassTable}.
\label{AgeMassFig}}
\end{figure}

\section{Hot Jupiter period distribution}
We take the opportunity to revisit the period distribution of gas giants in close orbits.  We have thus taken all planets with masses 0.15--12 M$_{\rm Jup}$ listed in TEPCat, and added the unpublished WASP planets as far as WASP-166b, and plot the cumulatative period distribution in Fig.~16.  This figure contains 321 planets out to 22 days, nearly doubling the 163 planets in the similar analysis in \citet{2012MNRAS.426..739H}.  

The two ``breaks'' suggested by \citet{2012MNRAS.426..739H} at 1.2 d and 2.7 d are still present.  The systems with periods $<$ 1.2 d are rare, despite having a greater range of inclinations that produce a transit, and despite being the easiest to find in transit surveys. They likely have short lifetimes owing to tidal inspiral.  Above 2.7 d the hot-Jupiter ``pileup'' continues to a more gradual rollover over the range 4--7 d.    Above $\sim$ 8 or 9 days the ground-based transit surveys will be less sensitive, and so one should be cautious in interpreting the distribution at longer periods. 

\begin{figure}
\mbox{\includegraphics[width=0.49\textwidth]{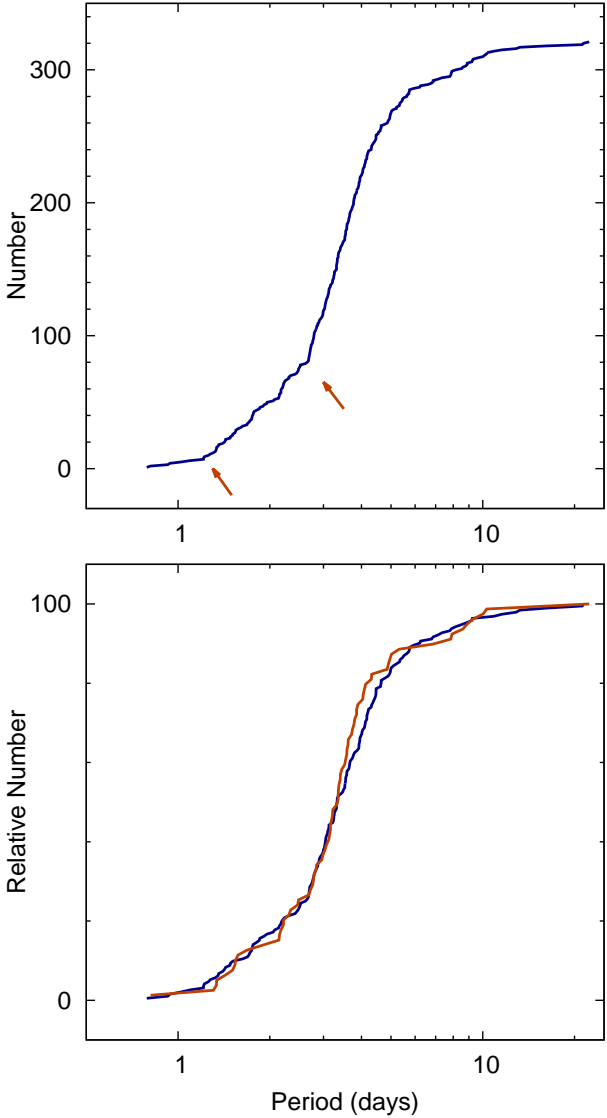}}
\caption{The cumulative period distribution for transiting planets in the range 0.15--12 M$_{\rm Jup}$. The red arrows mark breaks at 1.2 and 2.7 d suggested by Hellier \etal\ (2012). The lower panel compares the distributions for planets with host stars of above-solar (blue) and below-solar (red) metallicities.} 
\end{figure}

\citet{2013ApJ...767L..24D} analysed the {\it Kepler\/} sample of giant planets and found that the period distribution was strongly dependent on the metallicity of the host star (their Fig.~4).  They suggested that the hot-Jupiter bulge  is a feature only of metal-rich stars, and that the excess of hot Jupiters relative to longer-period giant planets is not present in a sample with [Fe/H] $<$ 0.    Note, however, that their analysis depended on the use of KIC metallicities, which come from photometric colours and thus may not be fully reliable \citep{2014ApJ...789L...3D}. 

Our above sample has very few planets beyond $P > 10$ d and so cannot be used to test the \citet{2013ApJ...767L..24D} result itself.  We can, however, address the related question of whether the period distribution {\it within\/} the hot-Jupiter bulge has a metallicity dependence, as might be the case if the formation of hot Jupiters depends strongly on metallicity.  We thus take all the planets in our sample with host-star metallicities listed in TEPCat, plus those in this paper, noting that  these metallicities come from spectroscopic analyses of relatively bright host stars. 

We then divide the sample into metallicities above and below solar (192 and 79 planets, respectively).  The two distributions are compared (after normalising them) in Fig.~16.   A K--S test says that they are not significantly different, with a 40\%\ chance of being drawn from the same distribution. Thus, there does not appear to be a metallicity dependence of the period distribution  {\it within\/} the hot-Jupiter bulge, though the discovery of more longer-period giant planets is needed to test the \citet{2013ApJ...767L..24D} result itself.

\section{Conclusions}
The ongoing WASP surveys continue to discover novel objects which push the bounds of known exoplanets (e.g.~the rapid circularisation timescale of WASP-140b) along with planets transiting bright stars which are good targets for atmospheric characterisation (e.g.~WASP-131b, with a $V$ = 10.1 star). We also present the longest-period (WASP-130b), lowest-mass (WASP-139b) and second-coolest (WASP-132b) of WASP-discovered planets.  We also demonstrate the power of WASP photometry in the possible detection of a 0.4-mmag rotational modulation of the star WASP-132.

\section*{Acknowledgements}
WASP-South is hosted by the South African Astronomical Observatory and
we are grateful for their ongoing support and assistance. Funding for
WASP comes from consortium universities and from the UK's Science and
Technology Facilities Council. The Euler Swiss telescope is supported
by the Swiss National Science Foundation. TRAPPIST is funded by the
Belgian Fund for Scientific Research (Fond National de la Recherche
Scientifique, FNRS) under the grant FRFC 2.5.594.09.F, with the
participation of the Swiss National Science Fundation (SNF). 
We acknowledge use of the ESO 3.6-m/HARPS under program 094.C-0090.

\bibliographystyle{mnras}
\bibliography{biblio}

\clearpage


\begin{table}
\renewcommand\thetable{A1}
\caption{Radial velocities.\protect\rule[-1.5mm]{0mm}{2mm}} 
\begin{tabular}{cccr} 
\hline 
BJD\,--\,2400\,000 & RV & $\sigma$$_{\rm RV}$ & Bisector \\
(UTC)  & (km s$^{-1}$) & (km s$^{-1}$) & (km s$^{-1}$)\\ [0.5mm] \hline
\multicolumn{4}{l}{{\bf WASP-130:}}\\  
56715.83348 & 1.3460 & 0.0065 & $-$0.0282\\
56718.84774 & 1.4386 & 0.0063 & $-$0.0140\\
56723.84644 & 1.5129 & 0.0072 & $-$0.0320\\
56725.73618 & 1.4049 & 0.0075 & $-$0.0361\\
56726.67716 & 1.3829 & 0.0076 & $-$0.0882\\
56773.77647 & 1.3618 & 0.0107 & $-$0.0173\\
56776.76305 & 1.4474 & 0.0067 &  0.0037\\
56778.76460 & 1.5612 & 0.0097 & $-$0.0311\\
56779.70963 & 1.5670 & 0.0065 & $-$0.0238\\
56803.66375 & 1.5488 & 0.0106 &  0.0160\\
56808.72754 & 1.3549 & 0.0095 & $-$0.0595\\
56810.69012 & 1.4195 & 0.0085 & $-$0.0280\\
56811.67572 & 1.4753 & 0.0072 &  0.0015\\
56833.65206 & 1.4397 & 0.0092 & $-$0.0070\\
56837.60754 & 1.5885 & 0.0100 & $-$0.0266\\
56853.59045 & 1.3982 & 0.0123 & $-$0.0281\\
56879.53529 & 1.4451 & 0.0132 & $-$0.0495\\ \cline{1-1} 
57017.84985 & 1.4195 & 0.0075 & $-$0.0481\\ 
57031.82387 & 1.5514 & 0.0078 & $-$0.0306\\ 
57044.82648 & 1.6247 & 0.0076 & 0.0136\\ 
57071.81936 & 1.4552 & 0.0076 & $-$0.0389\\ 
57082.85670 & 1.4949 & 0.0075 & $-$0.0088\\ 
57188.55692 & 1.4048 & 0.0112 & $-$0.0578\\ 
57189.66971 & 1.3880 & 0.0109 & $-$0.0189\\ 
57406.84526 & 1.4635 & 0.0090 & $-$0.0094\\ 
57426.81680 & 1.5852 & 0.0058 & $-$0.0316\\ 
57453.72114 & 1.4143 & 0.0051 & $-$0.0416\\ [0.5mm] 
\hline
\multicolumn{4}{l}{Bisector errors are twice RV errors} 
\end{tabular} 
\end{table}


\begin{tabular}{cccr} 
\hline 
BJD\,--\,2400\,000 & RV & $\sigma$$_{\rm RV}$ & Bisector \\
(UTC)  & (km s$^{-1}$) & (km s$^{-1}$) & (km s$^{-1}$)\\ [0.5mm] \hline
\multicolumn{4}{l}{{\bf WASP-131:}}\\  
56694.86288 & $-$19.6343 & 0.0055 & 0.0340\\ 
56713.72539 & $-$19.6948 & 0.0071 & 0.0276\\ 
56723.89584 & $-$19.6926 & 0.0063 & 0.0166\\ 
56724.87344 & $-$19.6800 & 0.0058 & 0.0354\\ 
56726.70668 & $-$19.6420 & 0.0057 & 0.0422\\ 
56744.91236 & $-$19.6899 & 0.0059 & 0.0347\\ 
56745.85830 & $-$19.6902 & 0.0061 & 0.0186\\ 
56746.62422 & $-$19.6705 & 0.0061 & 0.0149\\ 
56748.84647 & $-$19.6348 & 0.0059 & 0.0442\\ 
56749.72097 & $-$19.6652 & 0.0054 & 0.0173\\ 
56769.58299 & $-$19.6365 & 0.0056 & 0.0183\\ 
56809.72416 & $-$19.6991 & 0.0058 & 0.0322\\ 
56810.71410 & $-$19.6629 & 0.0072 & 0.0371\\ 
56811.72508 & $-$19.6332 & 0.0070 & 0.0233\\ 
56830.66400 & $-$19.6960 & 0.0064 & 0.0280\\ 
56839.62346 & $-$19.6526 & 0.0067 & 0.0218\\ 
56864.52557 & $-$19.6406 & 0.0056 & 0.0354\\ \cline{1-1} 
57055.82138 & $-$19.6531 & 0.0059 & 0.0170\\ 
57071.84521 & $-$19.6406 & 0.0065 & 0.0135\\ 
57110.88171 & $-$19.6367 & 0.0077 & 0.0145\\ 
57139.77294 & $-$19.6875 & 0.0085 & 0.0156\\ 
57194.63654 & $-$19.6207 & 0.0080 & 0.0252\\ 
57458.77449 & $-$19.6801 & 0.0055 & 0.0431\\ [0.5mm]
\multicolumn{4}{l}{{\bf WASP-132:}}\\  
56717.73665 & 31.1460 & 0.0103 & 0.0087\\
56749.88330 & 31.0358 & 0.0111 & 0.0298\\
56772.83833 & 31.0369 & 0.0099 & 0.0359\\
56779.80482 & 31.0630 & 0.0130 & $-$0.0239\\
56781.68823 & 31.1323 & 0.0116 & $-$0.0406\\
56782.75296 & 31.1388 & 0.0107 & 0.0058\\
56803.73759 & 31.0888 & 0.0232 & 0.0482\\
56808.75158 & 31.0784 & 0.0157 & 0.0360\\
56810.73750 & 31.1109 & 0.0153 & 0.0722\\
56811.69973 & 31.0846 & 0.0132 & 0.0369\\
56813.73159 & 31.0472 & 0.0281 & 0.0266\\
56814.74102 & 31.0197 & 0.0139 & $-$0.0298\\
56830.68796 & 31.0877 & 0.0141 & 0.0027\\
56833.58020 & 31.0764 & 0.0133 & 0.0523\\
56840.60592 & 31.0667 & 0.0196 & 0.0671\\
56853.53751 & 31.1467 & 0.0504 & 0.0238\\
56855.61273 & 31.0228 & 0.0127 & 0.0198\\
56856.59422 & 31.0223 & 0.0105 & 0.0341\\
56880.53254 & 31.0585 & 0.0151 & $-$0.0006\\
56888.54232 & 31.0888 & 0.0133 & 0.0140\\
56889.52621 & 31.0884 & 0.0110 & 0.0412\\
56910.48465 & 31.0966 & 0.0222 & 0.0143\\ \cline{1-1} 
57031.85994 & 31.0528 & 0.0150 & 0.0006\\
57072.87336 & 31.0326 & 0.0118 & $-$0.0502\\
57085.80107 & 30.9265 & 0.0128 & $-$0.0155\\
57086.74260 & 30.9483 & 0.0131 & $-$0.0098\\
57111.67974 & 30.9845 & 0.0173 & 0.0246\\
57112.85851 & 30.9697 & 0.0291 & $-$0.0224\\
57114.83854 & 30.9679 & 0.0212 & $-$0.0530\\
57139.68574 & 31.0283 & 0.0215 & $-$0.0124\\
57194.67667 & 31.0018 & 0.0278 & $-$0.0526\\
57405.83694 & 30.9468 & 0.0130 & $-$0.0141\\
57412.84526 & 30.9293 & 0.0176 & $-$0.0042\\
57426.84822 & 30.9744 & 0.0103 & 0.0005\\
57428.82227 & 30.9898 & 0.0084 & $-$0.0045\\
57455.88964 & 30.9471 & 0.0127 & $-$0.0609\\ [0.5mm] 
\hline
\multicolumn{4}{l}{Bisector errors are twice RV errors} 
\end{tabular} 



\begin{tabular}{cccr} 
\hline 
BJD\,--\,2400\,000 & RV & $\sigma$$_{\rm RV}$ & Bisector \\
(UTC)  & (km s$^{-1}$) & (km s$^{-1}$) & (km s$^{-1}$)\\ [0.5mm] \hline
\multicolumn{4}{l}{{\bf WASP-139:} CORALIE}\\  
54763.67767 & $-$13.0496 & 0.0201 & 0.0220 \\
54766.72371 & $-$13.0095 & 0.0157 & $-$0.0082 \\
54776.79057 & $-$12.9875 & 0.0254 & 0.0412 \\
56211.82564 & $-$12.9931 & 0.0143 & $-$0.0039 \\
56220.76495 & $-$13.0409 & 0.0129 & $-$0.0270 \\
56516.89696 & $-$13.0298 & 0.0145 & 0.0481 \\
56577.69538 & $-$13.0346 & 0.0114 & 0.0167 \\
56578.87291 & $-$12.9959 & 0.0105 & 0.0434 \\
56581.87217 & $-$13.0236 & 0.0101 & 0.0045 \\
56623.78855 & $-$13.0314 & 0.0111 & 0.0041 \\
56629.59554 & $-$13.0451 & 0.0116 & $-$0.0148 \\
56873.90932 & $-$13.0314 & 0.0189 & 0.0315 \\
56874.84054 & $-$12.9882 & 0.0137 & $-$0.0056 \\
56876.88875 & $-$13.0143 & 0.0148 & 0.0299 \\
56877.79318 & $-$13.0013 & 0.0168 & $-$0.0149 \\
56879.89171 & $-$12.9778 & 0.0143 & 0.0597 \\
56952.66033 & $-$12.9785 & 0.0170 & 0.0036 \\ \cline{1-1} 
56987.55377 & $-$12.9923 & 0.0413 & 0.0157 \\
56988.58979 & $-$12.9637 & 0.0170 & $-$0.0202 \\
57038.59542 & $-$12.9972 & 0.0184 & 0.0707 \\
57085.52067 & $-$13.0019 & 0.0154 & 0.0112 \\
57286.88716 & $-$13.0439 & 0.0340 & $-$0.0055 \\
57336.69857 & $-$13.0050 & 0.0302 & $-$0.0803 \\
57367.71629 & $-$12.9748 & 0.0168 & $-$0.0425 \\ [0.5mm]
\multicolumn{4}{l}{{\bf WASP-139:} HARPS}\\  
56927.79489 & $-$12.9888 & 0.0064 & $-$0.0186 \\
56929.83816 & $-$12.9931 & 0.0058 & $-$0.0043 \\
56948.67436 & $-$13.0052 & 0.0057 & 0.0002 \\
56949.65985 & $-$13.0067 & 0.0062 & 0.0179 \\
56951.74760 & $-$12.9814 & 0.0037 & 0.0021 \\
56952.74043 & $-$12.9857 & 0.0041 & $-$0.0181 \\
56953.76702 & $-$12.9860 & 0.0034 & 0.0121 \\
56955.70180 & $-$13.0079 & 0.0029 & 0.0066 \\
56957.86492 & $-$12.9971 & 0.0097 & $-$0.0048 \\
56958.72499 & $-$12.9856 & 0.0050 & 0.0157 \\
56959.70035 & $-$12.9962 & 0.0044 & 0.0032 \\
56959.71448 & $-$12.9984 & 0.0048 & 0.0080 \\
56959.72874 & $-$12.9900 & 0.0043 & 0.0126 \\
56959.74518 & $-$12.9994 & 0.0034 & 0.0066 \\
56959.76257 & $-$12.9952 & 0.0048 & 0.0093 \\
56959.78116 & $-$12.9929 & 0.0046 & 0.0014 \\
56959.79786 & $-$12.9955 & 0.0047 & 0.0101 \\
56959.81611 & $-$12.9907 & 0.0054 & $-$0.0037 \\
56959.83386 & $-$12.9970 & 0.0057 & 0.0112 \\
56959.85175 & $-$12.9952 & 0.0060 & $-$0.0198 \\
56959.86897 & $-$13.0020 & 0.0067 & 0.0217 \\
56997.70376 & $-$13.0057 & 0.0039 & 0.0043 \\
56999.72988 & $-$12.9820 & 0.0039 & 0.0058 \\
57032.61218 & $-$13.0151 & 0.0055 & $-$0.0028 \\
57033.58272 & $-$12.9941 & 0.0083 & 0.0073 \\
57034.60223 & $-$12.9789 & 0.0062 & 0.0048 \\
57035.58562 & $-$12.9898 & 0.0055 & 0.0182 \\ [0.5mm]
\hline
\multicolumn{4}{l}{Bisector errors are twice RV errors} 
\end{tabular}

\begin{tabular}{cccr} 
\hline 
BJD\,--\,2400\,000 & RV & $\sigma$$_{\rm RV}$ & Bisector \\
(UTC)  & (km s$^{-1}$) & (km s$^{-1}$) & (km s$^{-1}$)\\ [0.5mm] \hline
\multicolumn{4}{l}{{\bf WASP-140:}}\\  
56920.76077 & 2.5349 & 0.0098 &	0.0133	\\
56930.82132 & 1.7340 & 0.0078 & 0.0621 \\
56931.82218 & 2.5133 & 0.0072 & 0.0327 \\ 
56936.79830 & 2.3340 & 0.0080 & $-$0.0200 \\ 
56950.84894 & 1.7484 & 0.0075 & 0.0393 \\ 
56953.78131 & 2.1478 & 0.0094 & 0.0583 \\ 
56955.77226 & 1.8930 & 0.0080 & $-$0.0048 \\ 
56956.82473 & 2.4086 & 0.0108 & 0.0331 \\ 
56959.76870 & 1.7695 & 0.0083 & 0.0436 \\ 
56965.86872 & 2.3001 & 0.0075 & 0.0341 \\ \cline{1-1} 
56978.70617 & 2.5095 & 0.0120 & 0.0083 \\ 
56979.66055 & 1.9375 & 0.0122 & 0.0167 \\ 
56980.76045 & 2.3070 & 0.0108 & 0.0344 \\ 
56983.78948 & 2.2687 & 0.0101 & $-$0.0310 \\ 
56984.69862 & 1.7913 & 0.0094 & $-$0.0291 \\ 
56987.58015 & 2.4201 & 0.0165 & 0.0227 \\ 
57004.59553 & 1.7625 & 0.0100 & $-$0.0143 \\ 
57019.69293 & 2.1414 & 0.0129 & 0.0144 \\ 
57039.60065 & 2.3649 & 0.0105 & 0.0533 \\ 
57060.61575 & 1.7552 & 0.0100 & 0.0184 \\ 
57085.55015 & 1.9915 & 0.0094 & 0.0263 \\ 
57339.77164 & 1.8358 & 0.0124 & 0.0171 \\ 
57369.68260 & 2.1369 & 0.0230 & 0.0419 \\ [0.5mm] 
\hline
\multicolumn{4}{l}{Bisector errors are twice RV errors} 
\end{tabular} 

\begin{tabular}{cccr} 
\hline 
BJD\,--\,2400\,000 & RV & $\sigma$$_{\rm RV}$ & Bisector \\
(UTC)  & (km s$^{-1}$) & (km s$^{-1}$) & (km s$^{-1}$)\\ [0.5mm] \hline
\multicolumn{4}{l}{{\bf WASP-141:}}\\  
56955.84534 & 34.1669 & 0.0320 & 0.0300 \\ 
56965.81124 & 34.1291 & 0.0208 & $-$0.0239 \\ \cline{1-1}
56990.59092 & 33.5216 & 0.0406 & 0.0124 \\ 
56993.81477 & 33.5438 & 0.0472 & $-$0.0758 \\ 
57011.63059 & 33.9947 & 0.0268 & 0.0174 \\ 
57012.68022 & 33.9561 & 0.0401 & 0.0222 \\ 
57014.72766 & 33.8971 & 0.0300 & 0.0412 \\ 
57016.66733 & 33.6264 & 0.0293 & $-$0.0376 \\ 
57033.72918 & 33.4964 & 0.0310 & 0.0756 \\ 
57039.62726 & 33.7132 & 0.0392 & $-$0.0572 \\ 
57040.67281 & 33.5825 & 0.0376 & 0.0010 \\ 
57041.67337 & 34.1062 & 0.0497 & 0.0222 \\ 
57065.65032 & 34.0360 & 0.0361 & 0.0875 \\ 
57118.48484 & 34.0519 & 0.0695 & 0.0801 \\ 
57291.83917 & 33.5517 & 0.1031 & 0.1949 \\ 
57319.74513 & 34.0872 & 0.0708 & $-$0.1402 \\ 
57333.80278 & 33.9814 & 0.0534 & $-$0.0012 \\ 
57371.75720 & 33.6182 & 0.0412 & $-$0.0456 \\ [0.5mm] 
\multicolumn{4}{l}{{\bf WASP-142:}}\\  
56744.53717 & 47.2374 & 0.0201 & 0.0793 \\ 
56747.54086 & 47.0456 & 0.0262 & 0.0789 \\ 
56771.61011 & 47.1023 & 0.0260 & 0.0468 \\ 
56772.58229 & 47.0722 & 0.0258 & 0.0260 \\ 
56779.61261 & 47.2061 & 0.0321 & $-$0.0846 \\ 
56803.51322 & 47.1629 & 0.0277 & 0.0273 \\ \cline{1-1} 
56987.86072 & 46.9854 & 0.0584 & $-$0.1158 \\ 
56996.81308 & 47.1962 & 0.0423 & 0.0225 \\ 
57018.73174 & 47.0535 & 0.0390 & 0.0523 \\ 
57022.76169 & 46.9974 & 0.0330 & 0.0288 \\ 
57026.86587 & 46.9604 & 0.0300 & 0.0346 \\ 
57043.67305 & 47.1643 & 0.0568 & $-$0.0103 \\ 
57070.67363 & 47.2080 & 0.0326 & 0.0080 \\ 
57072.63982 & 47.2065 & 0.0425 & $-$0.0156 \\ 
57402.72844 & 47.0586 & 0.0499 & $-$0.0392 \\ 
57432.63093 & 47.1446 & 0.0269 & 0.0052 \\ 
\hline
\multicolumn{4}{l}{Bisector errors are twice RV errors} 
\end{tabular} 

\bsp	
\label{lastpage}
\end{document}